\RequirePackage[2018-12-01]{latexrelease}
\documentclass{rQUF2e}

\usepackage{epstopdf}
\usepackage{subfigure}
\usepackage{wrapfig}

\usepackage{xcolor}
\usepackage{booktabs}
\usepackage{array}
\usepackage{caption}
\usepackage{adjustbox}
\usepackage{siunitx}  
\usepackage{multirow}
\theoremstyle{plain}

\theoremstyle{definition}

\theoremstyle{remark}

\begin{document}

\title{Classifying and Clustering Trading Agents}

\author{Mateusz Wilinski$^{\ast}$$^\dag$\thanks{$^\ast$Corresponding author.
Email: mateusz.wilinski@tuni.fi}, Anubha Goel$^\dag$, Alexandros Iosifidis$^\ddag$ and Juho Kanniainen$^\dag$\\
\affil{
$^\dag$Financial Computing and Data Analytics, Tampere University, Finland\\
$^\ddag$Department of Electrical and Computer Engineering, Aarhus University, Denmark
}
\received{v1.0 released May 2025}}

\maketitle

\begin{abstract}
    The rapid development of sophisticated machine learning methods, together with the increased availability of financial data, has the potential to transform financial research, but also poses a challenge in terms of validation and interpretation.
    A good case study is the task of classifying financial investors based on their behavioral patterns.
    Not only do we have access to both classification and clustering tools for high-dimensional data, but also data identifying individual investors is finally available.
    The problem, however, is that we do not have access to ground truth when working with real-world data.
    This, together with often limited interpretability of modern machine learning methods, makes it difficult to fully utilize the available research potential.
    In order to deal with this challenge we propose to use a realistic agent-based model as a way to generate synthetic data.
    This way one has access to ground truth, large replicable data, and limitless research scenarios.
    Using this approach we show how, even when classifying trading agents in a supervised manner is relatively easy, a more realistic task of unsupervised clustering may give incorrect or even misleading results.
    We complete the results with investigating the details of how supervised techniques were able to successfully distinguish between different trading behaviors.
\end{abstract}


\section{Introduction}
Agent-based modeling (ABM) of the stock market has been an important research area for studying and predicting market dynamics through computational methods without relying on restrictive assumptions for analytical tractability \citep{farmer2009economy}.
In ABM, a finite number of agent classes drive market dynamics \citep{staccioli2021agent}, making the realistic characterization of these classes a central question.
However, two key limitations exist: (i) research data typically lack labels identifying agent categories, and (ii) the number and behavior of different agent types are often based on informed assumptions rather than actual investor-level observations.
Therefore, methods for linking investors to distinct behavioral categories based on data on agents’ actions are needed.
In this regard, \citet{kirilenko2017flash} categorized agents into high-frequency traders, market makers, fundamental buyers, fundamental sellers, and opportunistic traders based on transaction volume and scaled net positions.
More recently, \citet{cont2023analysis} applied an unsupervised spectral clustering algorithm to order flow data from a major broker to group traders into representative clusters with similar attributes.

However, a fundamental question has remained open: How accurately can investors be clustered or classified into the correct groups with investor-level limit order book data (level IV data)?
This question is crucial not only for ABM but also for the broader behavioral finance literature, where the role of different investor types, including HFT traders, fundamentalists, and noise traders, has been extensively studied \citep{grossman1988liquidity, shleifer1990noise, brogaard2014high, kalay2009detecting}.
Without sufficient data and a reliable method for assigning real investors to specific categories, empirical analyses of investor behavior risk becoming fragile or even misleading.

We address the question above by examining three key aspects: (i) available features, (ii) noise in individual trading activity, and (iii) machine learning methods applied to both unsupervised and supervised problems.
Regarding the first aspect, high-frequency trading data at the investor level is rarely available for research purposes, and existing datasets typically include only a limited set of features.\footnote{Some exceptions exist, including \citep{challet2018statistically}, which uses data from FX markets, \citep{cont2023analysis}, which includes investor-level order flow data from a major broker, and \citep{van2019high}, which uses proprietary data consisting of child order transactions from four large institutional investors. Moreover, \citep{brogaard2014high} and \citep{hagstromer2013diversity} use LOB data containing information about orders from high-frequency traders and market maker. For a more comprehensive literature review, see \citep{cont2023analysis}.}
It is therefore essential to ask whether the available features are sufficient to accurately link traders to their behavioral categories.
In this respect, the features used in \citep{cont2023analysis} are particularly interesting, as they are exceptionally rich and comprehensive.

The second aspect, concerning the magnitude of noise in individual trading activity, is important for several reasons.
First, categorizing investors into distinct groups is a highly idealized way of understanding investor behavior.
For example, some investors may provide liquidity in certain situations—behaving like market makers—while in other contexts, they may consume liquidity.
Secondly, even if an investor's overall strategy remains consistent with their assigned category, there may be situations in which various factors blur their behavioral profile.
For instance, so-called fundamentalists are not always fully informed and may, at times, behave similarly to noise traders.

The third aspect is very interesting for several reasons.
First, it is realistic to assume that researchers rarely have access to labeled data that would enable training machine learning classifiers via supervised learning.
Therefore, the performance of supervised classifiers is of interest because it provides an absolute upper bound for practical applications, where supervised learning is seldom feasible.
Comparing this with the use of unsupervised clustering methods -- such as those applied in \citep{cont2023analysis}, which do not use labels during training -- gives a more realistic sense of how well investors can actually be profiled.
Moreover, it is also relevant to examine how lighter-weight machine learning methods, such as support vector machines, differ in performance from deep learning approaches in this context.

To analyze the situation from all three perspectives, we developed a synthetic limit order book (LOB) environment using agent-based models.
Within this environment, we specified several investor types, including market makers, market takers, fundamentalists, chartists, and noise traders, totaling 15 investor categories.
This setup enabled us not only to utilize features previously available in empirical research but also to incorporate additional features that can improve clustering and classification accuracy.
Moreover, it allowed us to control the magnitude of noise within each investor category.
Most importantly, this approach gave us access to labeled data with ground truth, which enabled the use of supervised methods and allowed us to directly compare the performance of unsupervised clustering with supervised classification in this context.

The general idea of building an experimental synthetic environment has been utilized by many scientific groups currently working on applying ML in finance \citep{byrd2020abides, belcak2021fast}.
Such environments can be used to test capabilities and limitations of the ML methods in finance.
Especially when it comes to sophisticated techniques, which lack the potential to interpret and explain their results.
We hope this article will pave the way for more research focused on evaluating ML capabilities in helping financial researchers to answer questions where data is available, but ground truths are not.

Before we dig into details of our study, we would like to summarize our main findings.
First of all, we investigate the importance of feature selection.
We not only show which features are more important, but also how exactly do they allow to classify investors, and under what conditions.
Second, we investigate which types of investors are most difficult to group and how external knowledge can be used to reduce this difficulty.
Finally, we show that while both classification and clustering is possible with high accuracy, there is always a significant gap in favor of the former.

\section{Methods}

\subsection{Synthetic environment}

\begin{wrapfigure}{r}{0.5\textwidth}
\centering
\includegraphics[width=0.50\textwidth]{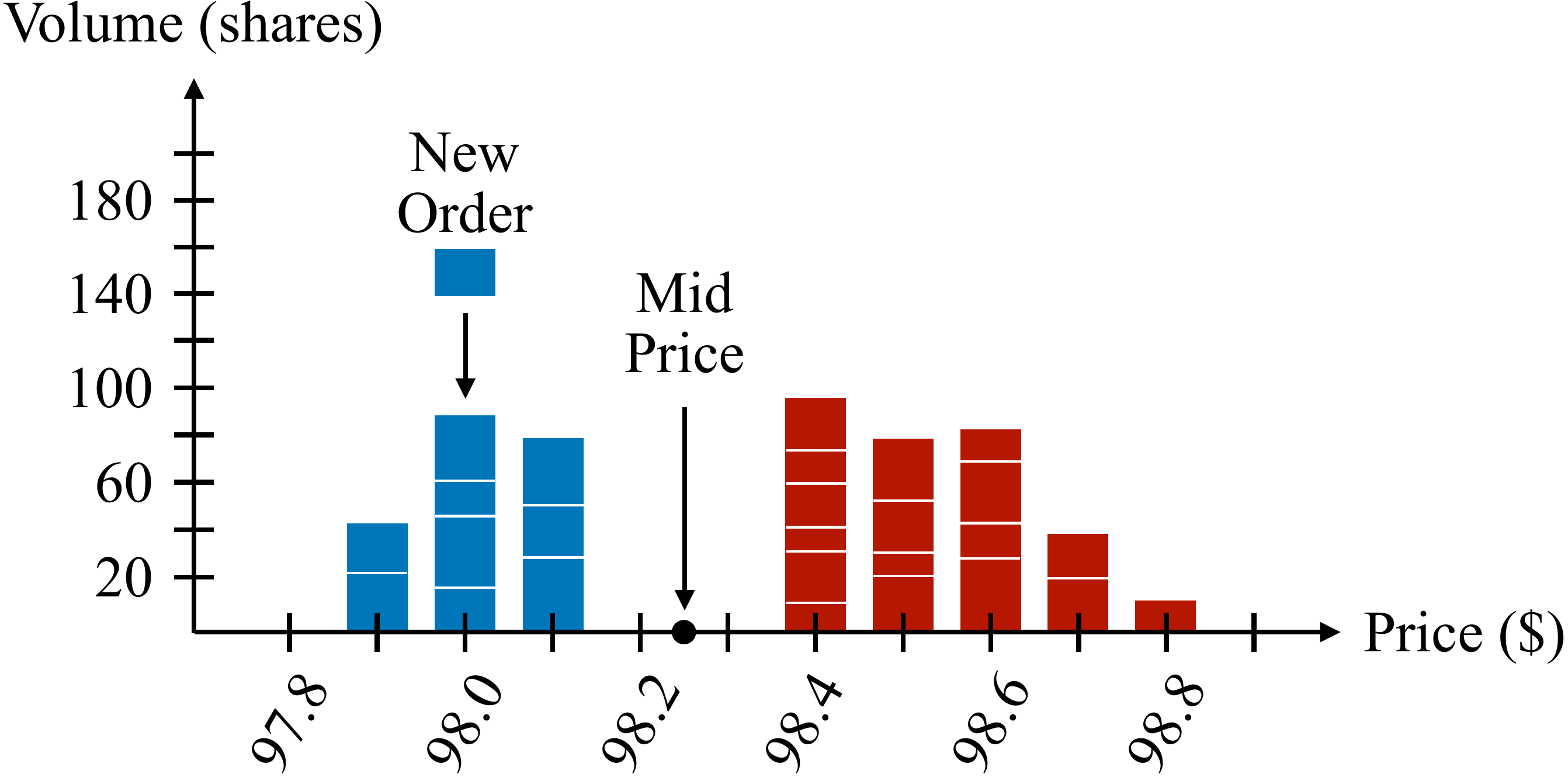}
\caption{A snapshot of a LOB with an incoming 20 shares buy order at a limit price of 98.00\$.
The best buy offer is at 98.10\$ and the best sell offer is at 98.40\$, which result in a 98.24\$ mid price and a 0.30\$ spread.
\label{fig:lob}}
\end{wrapfigure}

The data generated for the classification and clustering tasks comes from an artificial continuous double auction market.
In this market there are limit orders, with maximum or minimum execution price, and market orders, executed immediately with the best available price.
Existing, unfulfilled, orders are stored as a LOB \citep{gould2013limit} with two sides: ask side consisting of sell orders, and bid side consisting of buy orders.
Apart from sending an order, permitted operations also include modification and cancellation of an unfulfilled order.
There are four possible outcomes of any operation: order appearance, orders matching, order modification (volume change) and order deletion.
A single operation may result in one or multiple outcomes at the same time.
New orders are stored and matched using the price-time priority \citep{preis2011price}.
This means that higher price is prioritized for buy orders, while lower price is prioritized for sell orders.

An example of a LOB snapshot for a single asset is shown in Fig. \ref{fig:lob}.
There we can see nine unfulfilled buy orders, with a new arriving 20 shares limit order at a 98.00\$ price, and seventeen unfulfilled sell orders.
Together they create a slightly unbalanced (towards the ask side) setting with the best buy offer at 98.10\$ and the best sell offer at 98.40\$.
Hereafter, let us denote best buy (bid) offer as $b_t$, best sell (ask) offer as $a_t$, the mid-price as $p_t$ and the spread as $s_t$.
Note that $p_t = (a_t + b_t) / 2$ and $s_t = (a_t - b_t) / 2$.

The full implementation of the synthetic environment is available at \citep{wilinski2025simulator}, where all the technical details can be found.
The repository also includes every simulation setting with code needed to reproduce the results of this paper.
General concept of the LOB simulation is based on the idea of sending event messages rather than following discrete time steps.

\subsubsection{Agents}\label{sec:agents}

Actions in the simulations are taken by agents, which means that they are the only ones sending and modifying orders.
Each agent has a specific set of rules, according to which he trades.
There is no interaction between the agents, other than through the market.
Agents are not constrained with a budget and their behavior is not affected by any historical profits or losses.
We only consider a single asset simulation and agents' portfolio is limited to this asset.

Agents belong to a finite number of classes and the number of agents is significantly larger than the number of classes.
Here we propose five general classes of agents based on the typical classification, which can be found in the literature.
We further split them based on their parametrization into fifteen smaller classes covering all the artificial investors in our simulation.
In the simulation the sizes of all classes differ significantly, with noise trading dominating the market, as suggested in the literature \citep{delong1988survival}.
Below we give a detailed description of the five main classes of trading agents.
In Section \ref{sec:params}, we specify and justify the numbers and parameters used in the simulation.

The first group consists of \textbf{market makers}, whose main purpose is to provide liquidity by continuously placing buy and sell orders to earn the spread between bid and ask prices, facilitating smooth market operations.
Their role has been extensively studied, with early research highlighting their stabilizing effects \citep{grossman1988liquidity} and later studies examining their role in high-frequency trading \citep{menkveld2013high, brogaard2014high}.

Our definition of market makers strategy is following the description in \citep{chakraborty2011market, wah2017welfare}.
Each market maker appears in the market at exponentially distributed time intervals.
Once he arrives, he cancels all of his existing positions.
Then he creates a \textit{ladder} of limit orders at selling prices $A_t, A_t + q, \dots, A_t + K q$ and buying prices $B_t, B_t - q, \dots, B_t - K q$.
$A_t$ and $B_t$ are representing his view on respectively best-ask and best-bid orders, while $q$ and $K$ are representing the density and depth of his strategy.
In our simulations market makers assume that best-ask and best-bid orders are simply the best available offers, which means that $A_t = a_t$ and $B_t = b_t$.
Note that they only use limit orders and they do not modify them in between reconstructing their ladder.

The second group represents the \textbf{market takers} (liquidity traders), who try to execute large orders with as little impact as possible.
In contrast to liquidity providers, market takers operate within a specific time horizon, seeking to minimize trading costs, often for non-informational reasons such as portfolio re-balancing \citep[see, for example,][]{huberman2005optimal, kalay2009detecting}. 
In our simulation they simply cut the large order into smaller ones and then execute them in consecutive time intervals.
The size of the large order is fixed for each agent, but the smaller orders have a Gaussian distribution with fixed mean and standard deviation.
The time intervals between sending the smaller orders are also normally distributed with a fixed mean and standard deviation.
Both time intervals and small order sizes cannot go below 1.
After executing all the smaller chunks of a single large order, a market taker waits for an exponentially distributed period of time.
Then he needs to execute another large order.
Market takers use only market orders and do not engage in any actions other than the execution of large orders.

While previous groups were not reactive to the market conditions, \textbf{chartists} have a market dependent strategy.
They  rely on historical price patterns and technical analysis \citep{lo2000foundations}.
Advances in deep learning have expanded the applicability of technical analysis \citep{fischer2018deep}.
Our implementation of this group of investors is based on what was proposed in \citep{staccioli2021agent} with slight modifications.
Chartists believe that over a time horizon $h$, the price should change as:
\begin{equation}
    \hat{r}^C_{t+h} = w^C \cdot (p_t - p_{t-h}) + \epsilon_t,
\end{equation}
where $\hat{r}^C_{t+h}$ is the chartist's expectation of the price change in the time horizon $h$, $\epsilon_t$ is a Gaussian noise term with zero mean and a fixed standard deviation, and $w^C$ is a parameter.
Disregarding the noise, if $w^C$ is positive, chartists expect a continuation of the trend, while if it is negative, they expect a move in the opposite direction.
The former suggest a momentum strategy, while the latter should lead to something of a mean reverting strategy.
In our implementation a chartist arrives to the market to send a market order, which happens in exponentially distributed time intervals, or to send a limit order, which also happens in exponentially distributed time intervals.
In the first case, he checks whether his expected future price lands outside of the current spread.
If that is the case, he first cancels all of his inconsistent limit orders in the LOB and then he sends a marked order in line with his prediction.
In other words, if he expects the price to go up (down) beyond (below) the current best ask (bid) price, he cancels all of his own sell (buy) orders and then he sends a buy (sell) market order.
If his expected future price is inside of the spread, he does nothing.
In the second case, when he is interested in using a limit order, the chartist simply checks whether he expects the price to go up or down.
Then he cancels all of his existing inconsistent limit orders from the LOB and sends a new limit order in line with his expectation.
In this case inconsistent limit orders are the ones which are buy orders above the expected price and sell orders below the expected price.
The new limit order is a buy (sell) order with the expected price being the limit, assuming that this price is above (below) the current price.
All of the limit orders sent by chartists have an expiring period equal to the time horizon of their prediction.
After this period a limit order gets canceled.
The size of each order is drawn from a Gaussian distribution with fixed mean and standard deviation, but it is always at least one share.

While chartists were comparing the current price with the past, \textbf{fundamentalists} compare it with a hidden fundamental price, which is not known to others.
That is, fundamentalists base their investment decisions on their estimates of company value, trying to incorporate all relevant information on future dividends and discount factors.
When they observe deviations between market prices and their estimated `true' price, they trade accordingly.
This means that they expect the price to follow the fundamental value and they expect the future change to be equal to:
\begin{equation}
    \hat{r}^F_{t+h} = w^F \cdot (p_t^* - p_t) + \epsilon_t,
\end{equation}
where $\hat{r}^F_{t+h}$ is the fundamentalist's expectation of the price change in the time horizon $h$, $\epsilon_t$ is a Gaussian noise term with zero mean and a fixed standard deviation, and $w^F$ is a positive parameter.
This means that if the current price $p_t$ is below (above) the fundamental price $p_t^*$, he expects the price to go up (down) in the time horizon $h$.
Similarly to the chartists, fundamentalists have three types of actions: market orders, limit orders and cancellations.
The mechanism of these actions is exactly the same as for chartist.
The only difference is the way that the expected price change is computed.

The final group consists of \textbf{noise traders} whose trading strategy can be modeled as purely random from a point of view of an informed individual.
They make irrational investment decisions based on speculation rather than fundamental analysis \citep{shleifer1990noise}.
\citet{delong1990noise} find that prices can diverge significantly from fundamental values and that noise traders contribute to excess asset price volatility.
Our version of noise traders is capable of using both market and limit orders.
In both cases they send new orders in exponentially distributed time intervals.
In the case of market orders they simply randomly select the side (buy or sell with the same probability) and send the order to the LOB.
For limit orders they also pick the side randomly, but in addition they pick the price using a Gaussian distribution with mean equal to the current mid-price $p_t$ and a fixed standard deviation.
Note that the direction of their orders is disconnected from the chosen price.
The size of each order is drawn from a Gaussian distribution with fixed mean and standard deviation, but it is always at least one share.
Contrary to chartists or fundamentalists, they do not have a trading horizon.
Instead, they randomly cancel their orders.
Single cancellation for each noise trader appears with an exponential waiting time.

\subsubsection{Parametrization and stylized facts}\label{sec:params}

\begin{figure}
\begin{center}
\begin{minipage}{\textwidth}
\subfigure[one second price returns]{
\resizebox*{0.45\textwidth}{!}{\includegraphics{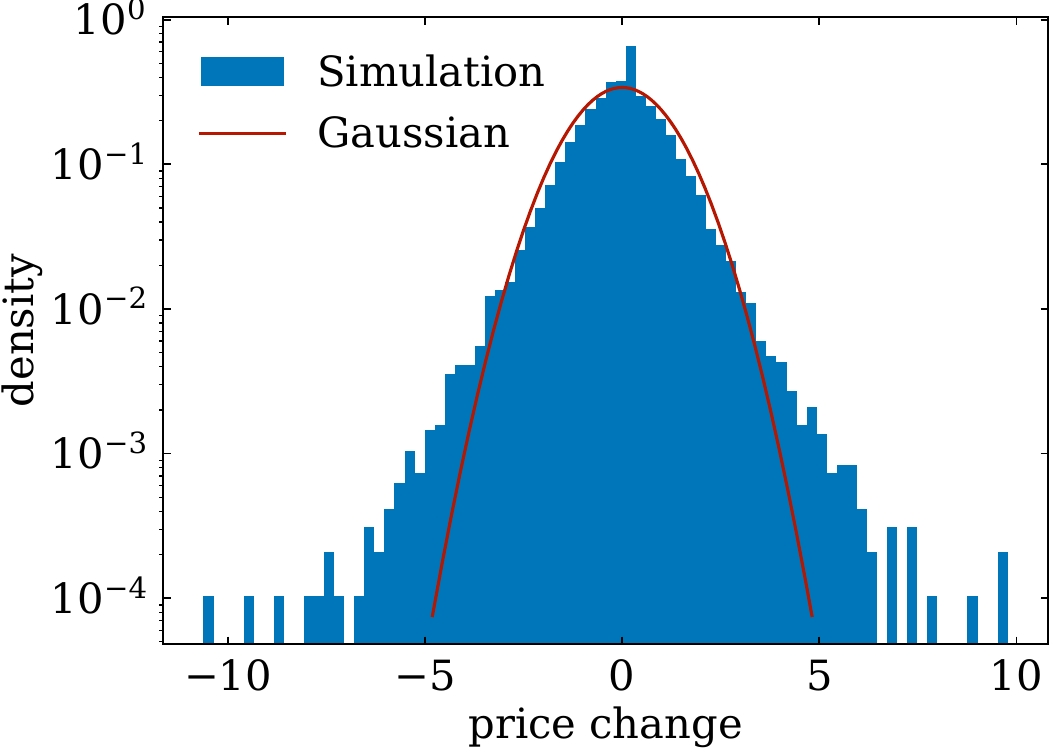}}\label{fig:sec_rets}}
\subfigure[one minute price returns]{
\resizebox*{0.45\textwidth}{!}{\includegraphics{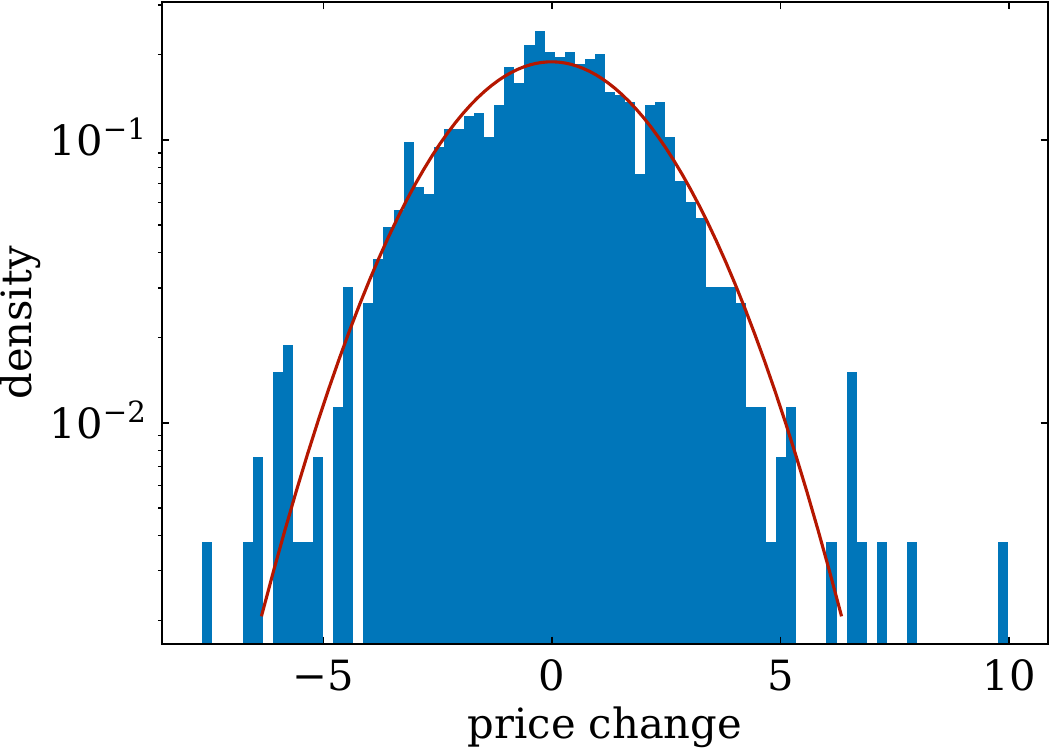}}\label{fig:min_rets}}
\caption{Histograms of price returns for two different time scales: (a) returns over one second; (b) returns over one minute.
Dark-red lines represent a Gaussian fit to the distributions.
For small time scales the distribution exhibits heavier tails with many non-Gaussian observations.
As time scale increases, we observe a more Gaussian distribution and less outliers.
\label{fig:returns}}
\end{minipage}
\end{center}
\end{figure}

The simulation is parametrized mainly through the number of specific agents and their parameters.
The way those numbers and parameters are set is constrained by a number of factors.
First of all we need to maintain sufficient liquidity to make sure that trading is continuous and realistic.
This requires enough activity on both market makers and noise traders.
Second, while we would like the price to be connected to some underlying fundamental price, we do want certain level of randomness, which requires to limit the influence of fundamental traders.
This also puts a boundary on the market makers, who otherwise would be able to keep the price at the same level for long periods of time.
Another aspect we took into consideration was that among agents with similar strategies, some would act in smaller time scales, while others would be less active.
This is why the main five classes of agents are further divided into fifteen classes.
Finally, we looked on how well our simulations track the known stylized facts observed in financial markets \citep{cont2001empirical}.
The latter guided the fractions of specific agents in the simulation and allowed to fine tune their strategies.
Similar approach was taken in \citep{vyetrenko2020get, staccioli2021agent}, where a number of statistical properties were computed for different settings of an ABM and then compared with real data.

While the goal of this article is not to create the most realistic model, we wanted to test at least the trademarks of true financial data.
As a result, we focused on the stylized facts related to the distribution of price returns and the autocorrelation structure of price returns.
As shown in Fig. \ref{fig:returns}, our simulation shows signs of two important properties of price returns distributions, observed in real data.
First, the tails of the distribution for higher frequencies are significantly wider than for the Gaussian fit of the distribution.
Second, as we decrease the frequency, the returns become more Gaussian.
When it comes to the autocorrelation structure, we looked at the raw autocorrelation of price returns, as well as the autocorrelation of absolute price returns.
The former is expected to be zero due to market efficiency.
While it is diminishing quickly, the simulation produces undesirable returns with immediate negative correlation.
Note that we are looking at the returns of the mid price, where the technical effect known as the bid-ask bounce \citep{abhyankar1997bid} should not appear.
Though this observation leaves space for improvement, it does not affect the main findings of our research.
As for the absolute returns, they show signs of long memory related to the volatility clustering observed in real data.

We did not focus so much on the raw activity, since the numbers depend highly on the liquidity of certain stocks and markets.
While some liquid stocks at Paris Bourse in 2001 had around 1300 transactions per hour \citep{bouchaud2002statistical}, 100 most traded stocks in Australia in 2000 had between 28 and 221 transactions per hour \citep{cao2009information}.
In \citep{makinen2019forecasting}, on the other hand, it was shown that for five liquid stocks at NASDAQ in 2014 there were between 1668 and 10878 transactions per hour.
In our experiments the number of transactions per hour is around 10000.

Building on our own experiments and suggestions from \citep{vyetrenko2020get, staccioli2021agent} we selected the following numbers of agents and values of their parameters.
We have three classes of market makers with different values of their parameters:
\begin{itemize}
    \item 20 agents with update rate $3000$, $K = 5$, $q = \frac{1}{4}$ and order size $5$.
    \item 20 agents with update rate $30000$, $K = 10$, $q = \frac{1}{2}$ and order size $5$.
    \item 20 agents with update rate $15000$, $K = 15$, $q = \frac{1}{4}$ and order size $5$.
\end{itemize}
Market takers are also divided into three groups:
\begin{itemize}
    \item 10 agents with large order rate $30000$, large order exit time $2000$ time units (subject to Gaussian noise with standard deviation $200$), large order size $100$ and average order chunk size $5$ with standard deviation $1.5$.
    \item 10 agents with large order rate $45000$, large order exit time $1000$ time units (subject to Gaussian noise with standard deviation $200$), large order size $400$ and average order chunk size $5$ with standard deviation $1.5$.
    \item 10 agents with large order rate $15000$, large order exit time $3000$ time units (subject to Gaussian noise with standard deviation $200$), large order size $50$ and average order chunk size $5$ with standard deviation $1.5$.
\end{itemize}
For agents reacting to market conditions, we divided chartist into four groups, including two trend followers and two with mean reverting view on the market:
\begin{itemize}
    \item 100 agents with $w^C_i = 1$, limit order rate $20000$, market order rate $10000$, noise standard deviation $0.1$, $10000$ time units horizon, average order size $5$ with standard deviation $1.5$.
    \item 100 agents with $w^C_i = \frac{1}{2}$, limit order rate $40000$, market order rate $20000$, noise standard deviation $0.1$, $40000$ time units horizon, average order size $5$ with standard deviation $1.5$.
    \item 100 agents with $w^C_i = -1$, limit order rate $20000$, market order rate $10000$, noise standard deviation $0.1$, $10000$ time units horizon, average order size $5$ with standard deviation $1.5$.
    \item 100 agents with $w^C_i = -\frac{1}{2}$, limit order rate $40000$, market order rate $20000$, noise standard deviation $0.1$, $40000$ time units horizon, average order size $5$ with standard deviation $1.5$.
\end{itemize}
The fundamentalists are divided into four groups, which differ on the expectation of how quickly the price will get back to its fundamental value:
\begin{itemize}
    \item 10 agents with $w^F_i = 1$, limit order rate $20000$, market order rate $10000$, noise standard deviation $0.1$, $10000$ time units horizon, average order size $5$ with standard deviation $1.5$.
    \item 10 agents with $w^F_i = \frac{1}{2}$, limit order rate $30000$, market order rate $20000$, noise standard deviation $0.1$, $20000$ time units horizon, average order size $5$ with standard deviation $1.5$.
    \item 10 agents with $w^F_i = \frac{1}{5}$, limit order rate $40000$, market order rate $20000$, noise standard deviation $0.1$, $40000$ time units horizon, average order size $5$ with standard deviation $1.5$.
    \item 10 agents with $w^F_i = \frac{1}{2}$, limit order rate $20000$, market order rate $10000$, noise standard deviation $0.1$, $40000$ time units horizon, average order size $5$ with standard deviation $1.5$.
\end{itemize}
Finally, noise traders form a single large group, similarly to the settings from \citep{vyetrenko2020get}:
\begin{itemize}
    \item 1060 agents with limit order rate $20000$, market order rate $10000$, cancellation rate $60000$, noise standard deviation $1.0$, average order size $5$ with standard deviation $1.5$.
\end{itemize}
The time unit of the simulation is $\frac{1}{10} \sec$ and each simulation consists of twenty hours of continuous trading activities.

An additional element of our simulation, which can be considered part of the parametrization, is the behavior of the fundamental price $p_t^*$.
This price is observed only by the fundamentalists.
All of them observe the same price, up to a noise factor.
In our simulation the fundamental price has the following dynamics:
\begin{equation}
    p_t^* =
    \begin{cases}
        100, & t \in (0, 180000) \cup (360000, 540000),\\
        70, & t \in (180000, 360000) \cup (540000, 720000).
    \end{cases}
\end{equation}
This means that during a single realization, fundamentalists observe three large jumps in the fundamental price.
On the one hand, we do not want the fundamental price to dominate the dynamics, hence the small number of fundamentalists and only three fundamental events per simulation.
On the other hand, this gives us an opportunity to leverage this information in search of a way to find the fundamentalists among other agents.

\begin{figure}
\begin{center}
\begin{minipage}{\textwidth}
\subfigure[price returns]{
\resizebox*{0.45\textwidth}{!}{\includegraphics{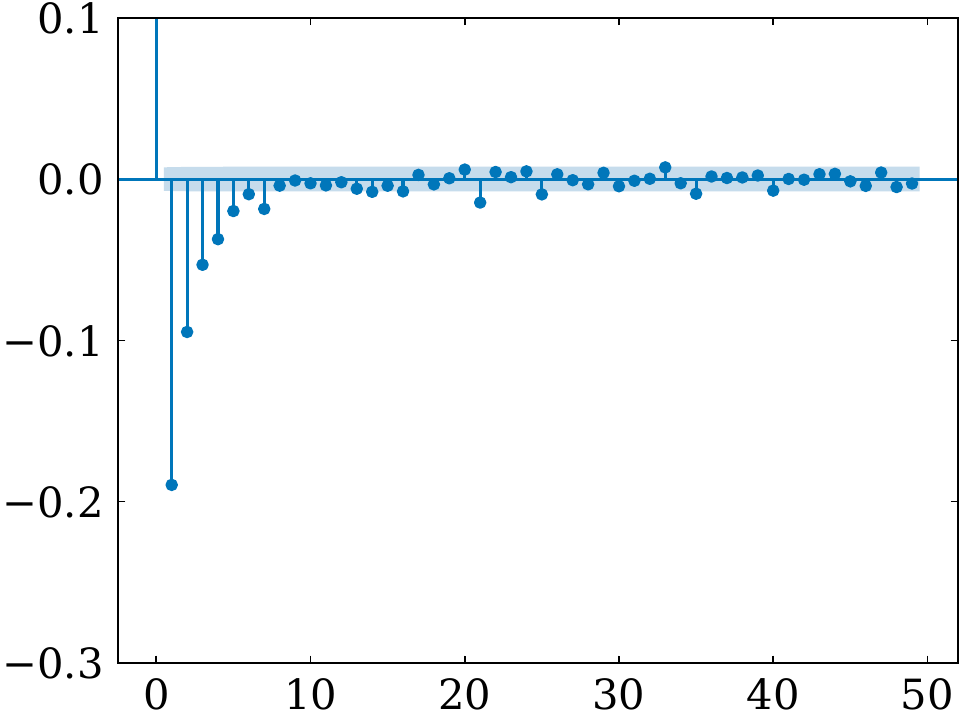}}\label{fig:price_corr}}
\subfigure[absolute price returns]{
\resizebox*{0.45\textwidth}{!}{\includegraphics{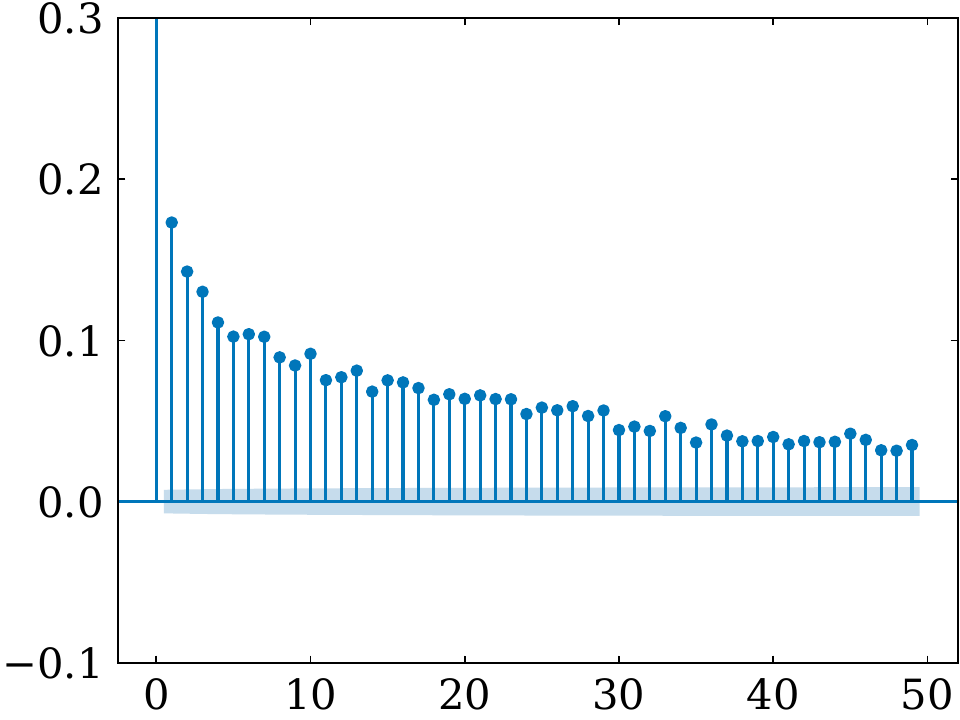}}\label{fig:abs_corr}}
\caption{Autocorrelation functions, with one second resolution, for (a) price returns and (b) absolute price returns.
\label{fig:corr}}
\end{minipage}
\end{center}
\end{figure}

\subsection{Features}\label{sec:features}

In order to both classify and cluster agents we propose the use of eighteen distinct features, which can be split into two groups.
We compute these features for each agent separately over all of their orders, trades, or cancellations.
First group consists of nine features used in the literature, mostly based on \citep{cont2023analysis} with some modifications in accordance with the nature of the simulation:
\begin{itemize}
    \item Buy ratio: percentage of buy orders among all orders.
    \item Cancellation ratio: percentage of orders, which are canceled before full execution.
    \item Number of trades: average number of trades per simulation run.
    \item Market ratio: percentage of market orders among all orders.
    \item Mean order creation time: average time between orders submission.
    \item Standard deviation of the creation time: standard deviation of the above.
    \item Mean order size: average order size.
    \item Standard deviation of the order size: standard deviation of the above.
    \item Total traded volume: sum of all the traded volume.
\end{itemize}
We propose additional nine features aiming at distinguishing certain behavior of the agents.
While they are partly based on our internal knowledge of the agents, we try to keep them as universal as possible:
\begin{itemize}
    \item Short trend: average absolute difference between the mid-price at the time an order was sent and the mid-price $10^3$ seconds before.
    \item Short directed trend: average product of the difference between the mid-price at the time an order was sent and the mid-price $10^3$ seconds before, and the order direction ($1$ for bid and $-1$ for ask).
    \item Medium trend: same as the short trend, but with the time horizon equal to $2 \cdot 10^3$ second instead of $10^3$ seconds.
    \item Medium directed trend: same as the short directed trend, but with the time horizon equal to $2 \cdot 10^3$ second instead of $10^3$ seconds.
    \item Long trend: same as the short trend, but with the time horizon equal to $4 \cdot 10^3$ second instead of $10^3$ seconds.
    \item Long directed trend: same as the short directed trend, but with the time horizon equal to $4 \cdot 10^3$ second instead of $10^3$ seconds.
    \item Fundamental profit: average profit around the fundamental events.
    Specifically, we take only the orders, which were send after the fundamental price change, but not more than $2 \cdot 10^3$ seconds after.
    For these we compute the current mid-price and the mid-price after $8 \cdot 10^3$ seconds.
    Then we use it to compute the mid-price return (over the above horizon), which we divide by the current mid price and multiply by the order direction ($1$ for bid and $-1$ for ask).
    This way we compute the order profit.
    \item Long fundamental profit: same as the fundamental profit, but the future mid-price is computed after $16 \cdot 10^3$ seconds.
    \item Weighted fundamental profit: same as the fundamental profit, but the average is weighted with the orders' sizes.
\end{itemize}
These extra features should allow us to distinguish momentum and mean reversion traders, as well as the fundamentalists, but without any knowledge of the actual dynamics of the fundamental price.

\subsection{Supervised classification methods}

For the purpose of classification we use two widely used classifiers, i.e., support vector machine (SVM) and deep neural networks (DNN).
Though we tested additional other methods, due to the consistency of the results and the length of the paper we decided to focus on these two methods.
This way we have one method representing classical ML approach and one from the modern family of neural architectures.
The two also have the advantage of having a certain level of interpretability.
For linear SVM, which was validated as the best SVM model, the interpretation is possible by comparing the magnitude of features' weights in this model \citep{guyon2003introduction}.
Note that, for multiple classes, it gives separate weights for each pair of classes as we use the one-vs-one scheme.
In other words, we can find which features are crucial in distinguishing between each pair of classes.
In the case of a DNN, we utilize layer-wise relevance propagation \citep{montavon2019layer}.
This method shows how strongly different features affected the selection of a given class for a specific data point.
This allows to find the statistically crucial features for each class of agents.

\subsection{Unsupervised clustering methods}

In our initial search for the most suited method we tested five methods: K-means, K-medoids, hierarchical clustering, spectral clustering, and kernel K-means.
This selection represents a broad spectrum of clustering paradigms, with K-means, K-medoids, and Agglomerative hierarchical clustering serving as representatives of classical linear approaches, while spectral clustering and kernel K-means capture non-linear structures.
Given the challenge of model selection in clustering, we adopted a stability-based evaluation criterion proposed in \citep{mourer2023selecting}.
This method balances between-cluster and within-cluster stability, ensuring that a robust clustering solution exhibits high stability across repeated perturbations while avoiding the presence of stable sub-structures within individual clusters.

Due to the computational complexity associated with spectral and kernel K-means, we had to restrict the clustering model evaluation to a random subset of five simulations.
For each method, stability scores were computed for the number of clusters varying from  $k=1$ to $k=16$.
Among the tested approaches, hierarchical clustering consistently exhibited high and stable values of stability scores.
This suggests that hierarchical clustering effectively captures the inherent structure of the data while remaining robust to perturbations and noise.
Therefore, and due to computational complexity limitations of other methods, we adopt hierarchical clustering as our clustering method.

\section{Results}

Using the parametrization described in Sec. \ref{sec:params} we run forty trading simulations, each twenty-hours long.
We then use the generated data to produce three settings representing different levels of noise.
In the first version, we compute the features described in Sec. \ref{sec:features} for each of the 1590 agents.
Though the behavior of each agent is subject to certain levels of randomness, as described in Sec. \ref{sec:agents}, we consider this setting as no additional noise.
In the second setting, we take 530 out of 1060 noise traders and we merge their actions with 530 other agents.
In other words, activities of each agent not being a noise trader (market maker, market taker, chartist or fundamentalist) are joined with activities of a single noise traders and they are treat as coming from a single agent.
This way we end up with 1060 agents, out of which 530 are noise traders, while other groups remain the same size.
We consider this setting to be a 50\% noise setting, even though the additional noise actions may not necessary constitute 50\% of given agent's actions.
The third setting is similar to the second one, but now each non-noise agent is merged with two, instead of one, noise traders.
We refer to this setting as a 66.6\% noise setting.
Though we understand the limitations of such setting, including the different way it affects different agent classes makes the results much more comparable.
This way, market conditions and activities are literally the same, which would be very difficult to obtain differently.

All of the above settings are tested in two scenarios.
In the first one we use all features described in Sec. \ref{sec:features}, including the ones handcrafted by us.
The second scenario, on the other hand, includes only the first nine features described in Sec. \ref{sec:features}.
This way, we want to test how successful are the basic features used in literature, and how much performance can be improved by adopting a more informed approach.
Obviously, the latter is not always available, but we believe it is still valuable to evaluate the difference between the two approaches.
All of the presented results can be reproduced using the code available at \citep{wilinski2025classifying}.
One can also find all the technical details there.

\subsection{Classification}

For classification, we divide the data into training, validation, and test sets.
Each of them have the same fractions of different agent types as the full data and they are disjoint.
The training set consists of 60\% of all agents, the validation set consists of 10\% of all agents and test dataset consists of the remaining 30\%.
The training set is used to train the classification models, while the validation set is used to fine-tune hyper-parameters of the models.
All the presented results are obtained by measuring the performance of the best models on the test set.

\begin{figure}
\begin{center}
\begin{minipage}{\textwidth}
\subfigure[Scenario with 18 features and no noise. Overall accuracy 0.99.]{
\resizebox*{0.32\textwidth}{!}{\includegraphics{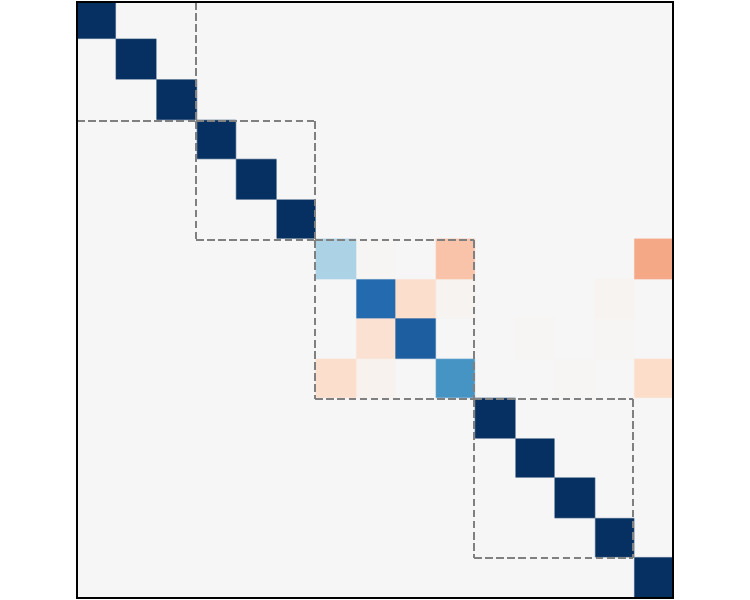}}\label{fig:svm_cm_1_18}}
\subfigure[Scenario with 18 features and 50\% noise. Overall accuracy 0.96.]{
\resizebox*{0.32\textwidth}{!}{\includegraphics{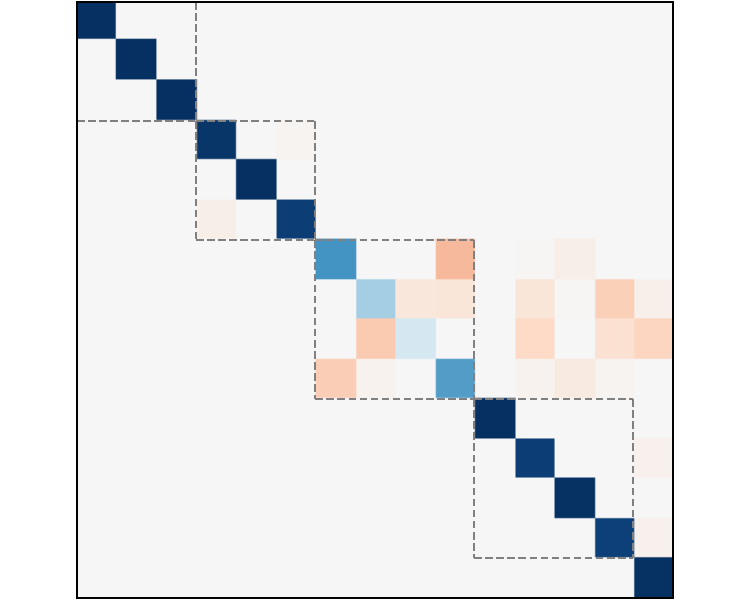}}\label{fig:svm_cm_2_18}}
\subfigure[Scenario with 18 features and 66.6\% noise. Overall accuracy 0.91.]{
\resizebox*{0.32\textwidth}{!}{\includegraphics{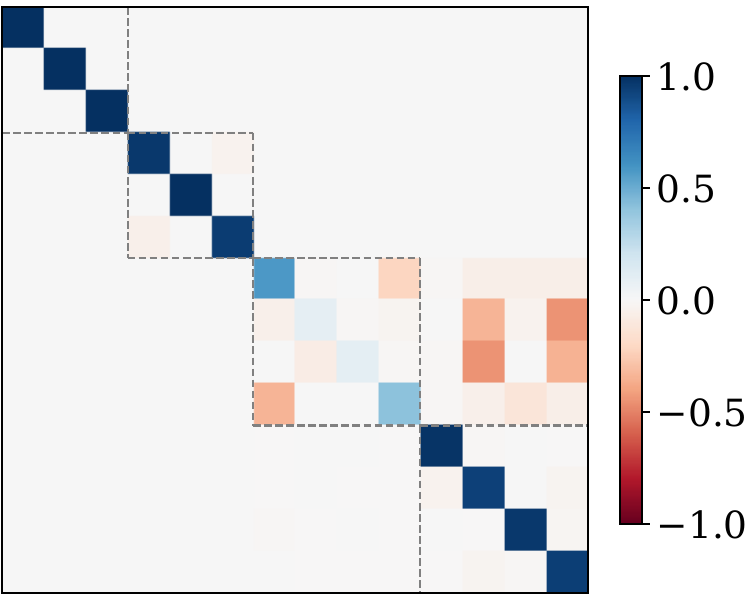}}\label{fig:svm_cm_3_18}}
\subfigure[Scenario with 9 features and no noise. Overall accuracy 0.85.]{
\resizebox*{0.32\textwidth}{!}{\includegraphics{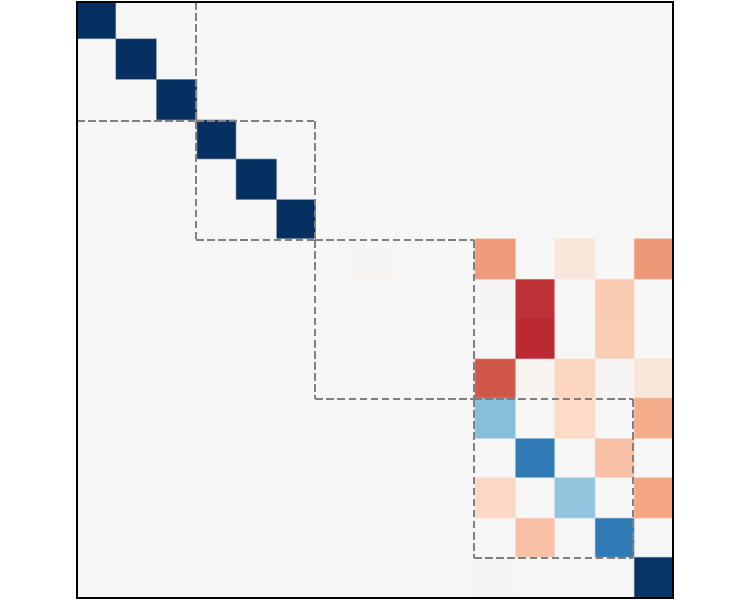}}\label{fig:svm_cm_1_9}}
\subfigure[Scenario with 9 features and 50\% noise. Overall accuracy 0.79.]{
\resizebox*{0.32\textwidth}{!}{\includegraphics{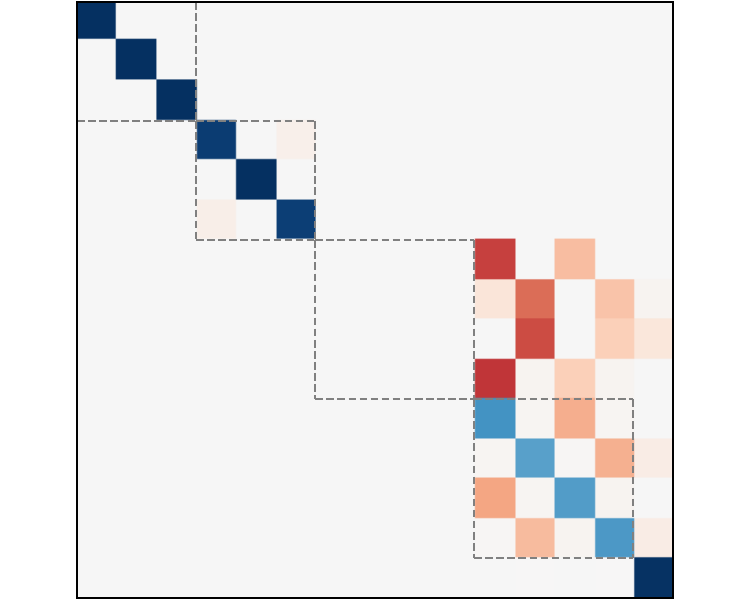}}\label{fig:svm_cm_2_9}}
\subfigure[Scenario with 9 features and 66.6\% noise. Overall accuracy 0.58.]{
\resizebox*{0.32\textwidth}{!}{\includegraphics{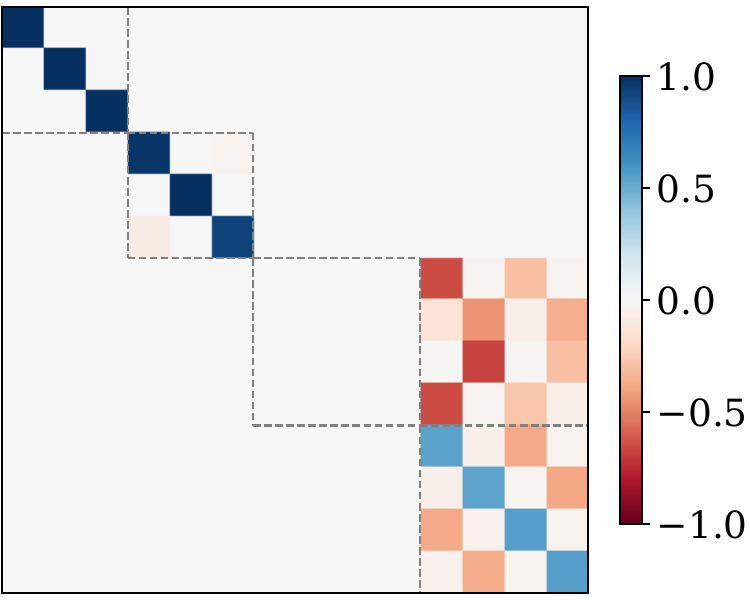}}\label{fig:svm_cm_3_9}}
\caption{Confusion matrix for classifying trading investors using SVM for different settings.
Dashed lines separate internal matrices for market makers (top left sub-matrix), market takers (sub-matrix second from the top), fundamentalists (sub-matrix third from the top) and chartists (down right sub-matrix).
\label{fig:svm_cm}}
\end{minipage}
\end{center}
\end{figure}

\begin{figure}
\begin{center}
\begin{minipage}{\textwidth}
\subfigure[Scenario with 18 features and no noise. Overall accuracy 0.99.]{
\resizebox*{0.32\textwidth}{!}{\includegraphics{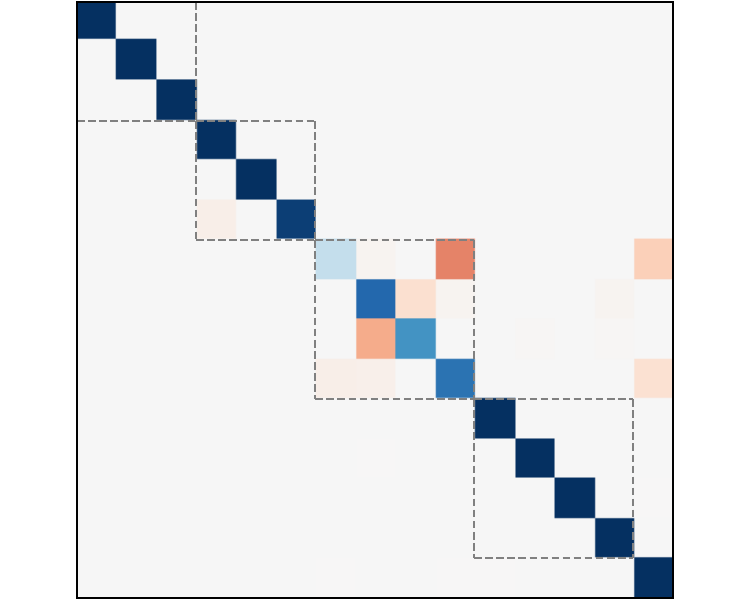}}\label{fig:dnn_cm_1_18}}
\subfigure[Scenario with 18 features and 50\% noise. Overall accuracy 0.95.]{
\resizebox*{0.32\textwidth}{!}{\includegraphics{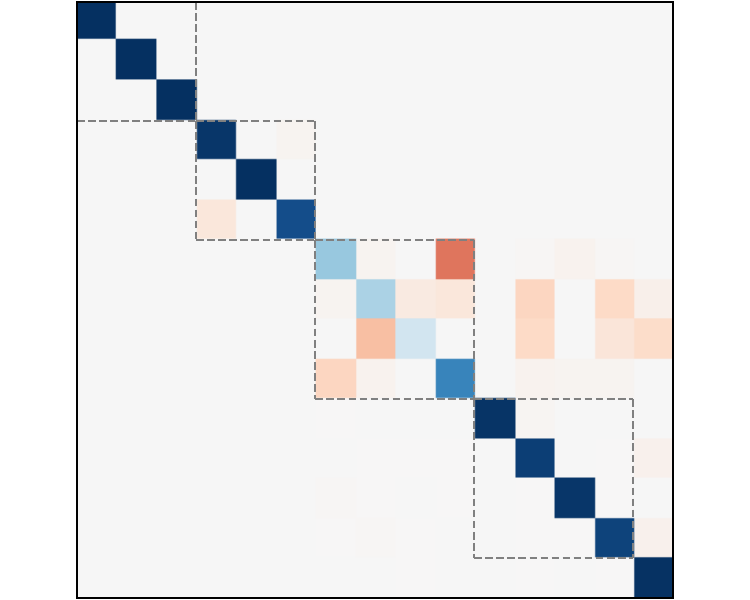}}\label{fig:dnn_cm_2_18}}
\subfigure[Scenario with 18 features and 66.6\% noise. Overall accuracy 0.90.]{
\resizebox*{0.32\textwidth}{!}{\includegraphics{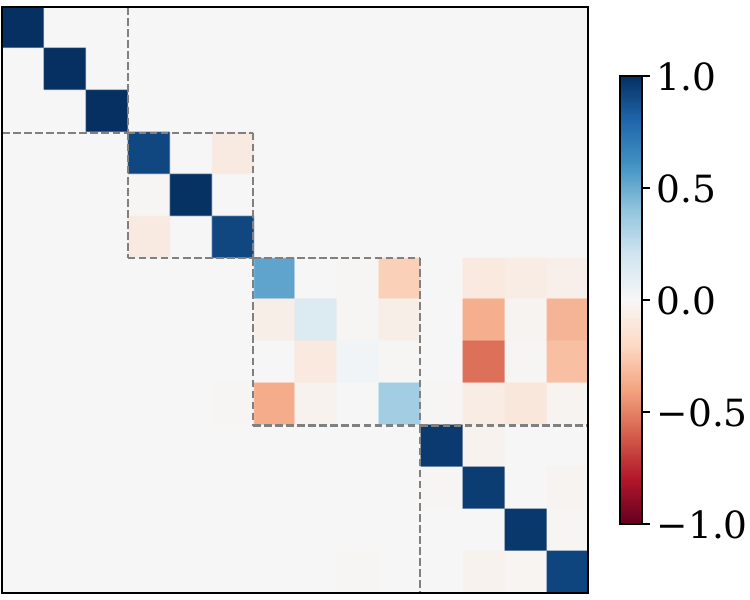}}\label{fig:dnn_cm_3_18}}
\subfigure[Scenario with 9 features and no noise. Overall accuracy 0.85.]{
\resizebox*{0.32\textwidth}{!}{\includegraphics{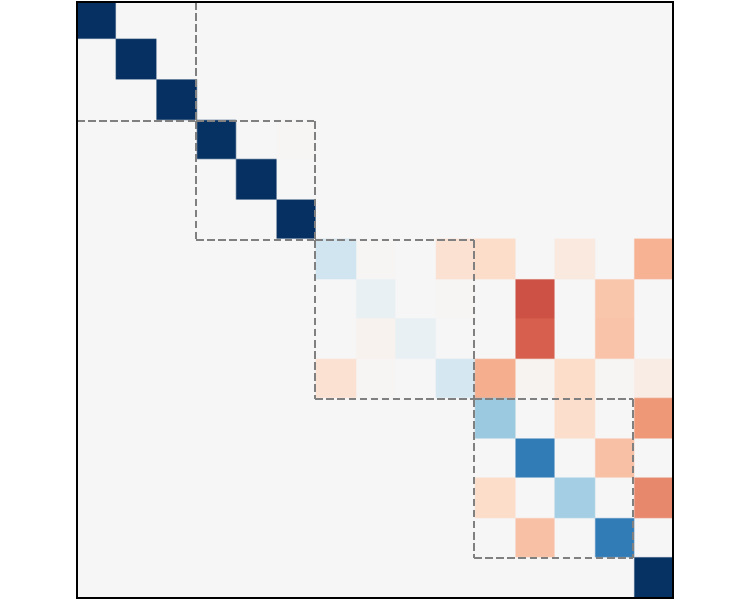}}\label{fig:dnn_cm_1_9}}
\subfigure[Scenario with 9 features and 50\% noise. Overall accuracy 0.79.]{
\resizebox*{0.32\textwidth}{!}{\includegraphics{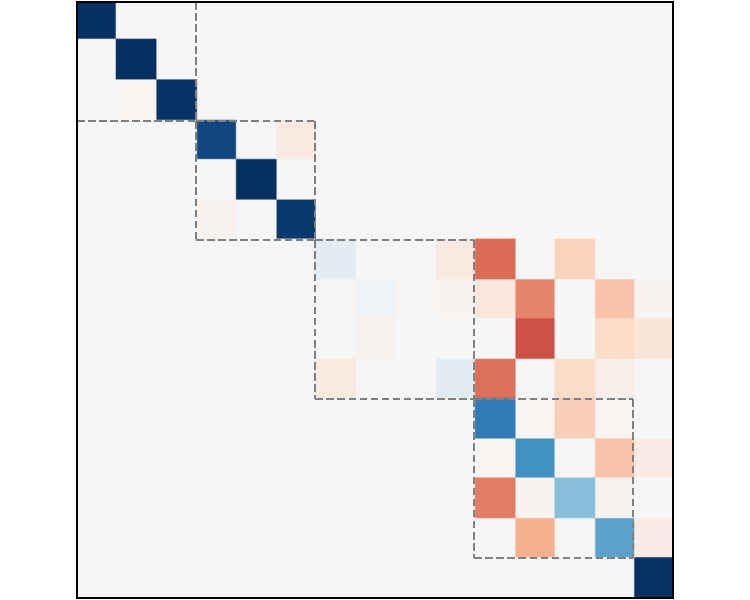}}\label{fig:dnn_cm_2_9}}
\subfigure[Scenario with 9 features and 66.6\% noise. Overall accuracy 0.57.]{
\resizebox*{0.32\textwidth}{!}{\includegraphics{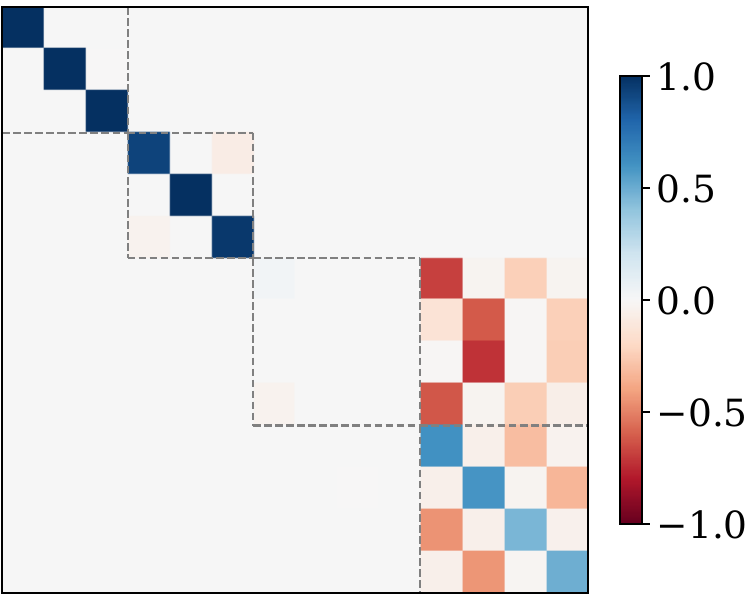}}\label{fig:dnn_cm_3_9}}
\caption{Confusion matrix for classifying trading investors using DNN architecture for different settings.
Dashed lines separate internal matrices for market makers (top left sub-matrix), market takers (sub-matrix second from the top), fundamentalists (sub-matrix third from the top) and chartists (down right sub-matrix).
\label{fig:dnn_cm}}
\end{minipage}
\end{center}
\end{figure}

In order to find the best SVM hyper-parameters, which include the kernel type and its hyper-parameter values, we performed grid-search based on the performance of the trained models on the validation set.
We tested linear, polynomial and radial basis function kernels.
In all cases we tested the regularization parameter $C$ values of 1, 10, 100 and 1000.
For the radial basis function kernel and the polynomial kernel we tested the $\gamma$ parameter values of 0.01, 0.1 and 1.
For the polynomial kernel we additionally tested the degree values of 2, 3 and 4.
The linear kernel with regularization parameter equal to one was found to provide the best performance for both 9 and 18 features settings.
While the linear kernel was consistently better then others, the change of the regularization parameter had a very small effect on the results.
For the DNN architecture we used the Ray Tune library \citep{liaw2018tune} to determine the number and sizes of the hidden layers, the batch size, the dropout percentage, and the learning rate hyper-parameter values.
We assumed a structure with four or five layers, i.e., two or three hidden layers with sizes being powers of two ranging from $2^6$, till $2^11$.
Each layer, apart from the last one, had the rectified linear unit activation function and was subject to dropout (see \citep{liu2017survey} for more details), with a dropout rate selected from the values 0.1, 0.2, or 0.3.
The tested batch sizes were consecutive powers of two between $2^3$ and $2^6$, and the learning rate was selected from a log-uniform distribution between $10^{-4}$ and $10^{-2}$.
The chosen architecture was formed by three hidden layers with 256, 1024, and 1024 neurons, respectively, dropout rate of 0.2, batch size equal to 32, and learning rate equal to $10^{-3}$.
We noticed that many different, often distant, sets of DNN parameters led to a similar performance to that obtained by the chosen architecture.

\begin{table}
\begin{center}
\begin{minipage}{80mm}
\tbl{Classification Report for SVM architecture and 18 features.}
{\begin{tabular}{@{}c c c c c c c c c}\toprule
 & \multicolumn{4}{c}{\textbf{no-noise}} & \multicolumn{4}{c}{\textbf{66.6\% noise}} \\
\colrule
\textbf{Class} & \textbf{Precision} & \textbf{Recall} & \textbf{F1-score} & \textbf{Support} & \textbf{Precision} & \textbf{Recall} & \textbf{F1-score} & \textbf{Support} \\
\colrule
market maker(1) & 1.00 & 1.00 & 1.00 & 160 & 1.00 & 1.00 & 1.00 & 160 \\
market maker(2) & 1.00 & 1.00 & 1.00 & 160 & 1.00 & 1.00 & 1.00 & 160 \\
market maker(3) & 1.00 & 1.00 & 1.00 & 160 & 1.00 & 1.00 & 1.00 & 160 \\
market taker(1) & 1.00 & 1.00 & 1.00 & 80 & 0.95 & 0.96 & 0.96 & 80 \\
market taker(2) & 1.00 & 1.00 & 1.00 & 80 & 1.00 & 1.00 & 1.00 & 80 \\
market taker(3) & 1.00 & 1.00 & 1.00 & 80 & 0.96 & 0.95 & 0.96 & 80 \\
fundamentalist(1) & 0.60 & 0.31 & 0.41 & 80 & 0.53 & 0.57 & 0.55 & 80 \\
fundamentalist(2) & 0.78 & 0.78 & 0.78 & 80 & 0.41 & 0.09 & 0.14 & 80 \\
fundamentalist(3) & 0.82 & 0.82 & 0.82 & 80 & 0.57 & 0.10 & 0.17 & 80 \\
fundamentalist(4) & 0.61 & 0.59 & 0.60 & 80 & 0.52 & 0.41 & 0.46 & 80 \\
chartist(1) & 1.00 & 1.00 & 1.00 & 800 & 0.96 & 0.98 & 0.97 & 800 \\
chartist(2) & 1.00 & 1.00 & 1.00 & 800 & 0.87 & 0.93 & 0.90 & 800 \\
chartist(3) & 1.00 & 1.00 & 1.00 & 800 & 0.96 & 0.97 & 0.97 & 800 \\
chartist(4) & 1.00 & 1.00 & 1.00 & 800 & 0.87 & 0.94 & 0.91 & 800 \\
noise trader(1) & 0.99 & 1.00 & 1.00 & 8480 & N/A & N/A & N/A & N/A \\
\textbf{Overall Accuracy} & \multicolumn{4}{c}{\textbf{0.99}} & \multicolumn{4}{c}{\textbf{0.91}} \\ 
\botrule
\end{tabular}}
\label{tab:svm_report_18}
\end{minipage}
\end{center}
\end{table}

We report the results of classification using confusion matrices, plots, and tables describing detailed information about precision, recall and F1-score \citep{powers2011evaluation}.
Confusion matrices for both SVM and DNN, in all 6 scenarios with different numbers of features and different fraction of noise, are available respectively in Fig. \ref{fig:svm_cm} and Fig. \ref{fig:dnn_cm}.
These matrices represent the fractions of falsely and correctly classified agents.
For readability we made a little change in the way the confusion matrix is typically represented, by making all the off-diagonal values negative.
This way non-zero values on the diagonal are immediately seen as positive, while non-zero off-diagonal values are immediately recognized as misclassifications.
To be more precise, diagonal values are between 0 and 1, and they represent the fraction of given class' agents correctly classified by the classifier.
On the other hand, off-diagonal values are between -1 and 0, and when the value $c_{ij}$ in the $i$th row and the $j$th column is smaller than zero, then $-c_{ij}$ fraction of agents from class $i$ was classified as belonging to class $j$.
The order of rows and columns is the same as in Sec. \ref{sec:agents}, and additionally we used dashed lines to show internal confusion matrices for market makers, market takers, fundamentalists and chartists.
The latter makes it easier to see which classes pose more difficulties and what are the misclassification trends.
When reporting precision, recall and F1-score, we only show the no-noise and the 66.6\% noise scenarios, since the trends are obvious and we do not want to extend the already lengthy manuscript.

\begin{table}
\begin{center}
\begin{minipage}{80mm}
\tbl{Classification Report for SVM and 9 features.}
{\begin{tabular}{@{}c c c c c c c c c}\toprule
 & \multicolumn{4}{c}{\textbf{no-noise}} & \multicolumn{4}{c}{\textbf{66.6\% noise}} \\
\colrule
\textbf{Class} & \textbf{Precision} & \textbf{Recall} & \textbf{F1-score} & \textbf{Support} & \textbf{Precision} & \textbf{Recall} & \textbf{F1-score} & \textbf{Support} \\
\colrule
market maker(1) & 1.00 & 1.00 & 1.00 & 160 & 1.00 & 1.00 & 1.00 & 160 \\
market maker(2) & 1.00 & 1.00 & 1.00 & 160 & 1.00 & 1.00 & 1.00 & 160 \\
market maker(3) & 1.00 & 1.00 & 1.00 & 160 & 1.00 & 1.00 & 1.00 & 160 \\
market taker(1) & 1.00 & 1.00 & 1.00 & 80 & 0.93 & 0.97 & 0.95 & 80 \\
market taker(2) & 1.00 & 1.00 & 1.00 & 80 & 1.00 & 1.00 & 1.00 & 80 \\
market taker(3) & 1.00 & 1.00 & 1.00 & 80 & 0.97 & 0.93 & 0.95 & 80 \\
fundamentalist(1) & 0.00 & 0.00 & 0.00 & 80 & 0.00 & 0.00 & 0.00 & 80 \\
fundamentalist(2) & 0.00 & 0.00 & 0.00 & 80 & 0.00 & 0.00 & 0.00 & 80 \\
fundamentalist(3) & 0.00 & 0.00 & 0.00 & 80 & 0.00 & 0.00 & 0.00 & 80 \\
fundamentalist(4) & 0.00 & 0.00 & 0.00 & 80 & 0.00 & 0.00 & 0.00 & 80 \\
chartist(1) & 0.51 & 0.42 & 0.46 & 800 & 0.47 & 0.54 & 0.50 & 800 \\
chartist(2) & 0.61 & 0.70 & 0.66 & 800 & 0.48 & 0.53 & 0.50 & 800 \\
chartist(3) & 0.56 & 0.40 & 0.46 & 800 & 0.52 & 0.55 & 0.53 & 800 \\
chartist(4) & 0.67 & 0.70 & 0.69 & 800 & 0.51 & 0.55 & 0.53 & 800 \\
noise trader(1) & 0.93 & 0.98 & 0.95 & 8480 & N/A & N/A & N/A & N/A \\
\textbf{Overall Accuracy} & \multicolumn{4}{c}{\textbf{0.85}} & \multicolumn{4}{c}{\textbf{0.58}} \\ 
\botrule
\end{tabular}}
\label{tab:svm_report_9}
\end{minipage}
\end{center}
\end{table}

The first observation is that the market makers, in the current form, are easily classified regardless of the size of the noise and the features used.
It is confirmed by both methods in all used scenarios.
While these agents have very specific strategies, it is interesting that noise levels have practically no effect on the results here.
Market takers, who are very different from other agents, are also relatively easy to classify.
However, we observed that in this case noise plays some role and SVM gives slightly better results when it happens.
As expected, fundamentalists are most difficult to classify.
Even when we all features are used in the no-noise scenario, both SVM and DNN struggle to classify the agents correctly, but the misclassification is mostly among different types of fundamentalists and noise traders.
As the noise increases, fundamentalists become more often misclassified as chartists.
When fewer features are used, we see the main difference between the results obtained with SVM and DNN.
While SVM classifies all the fundamentalists as chartists and noise traders in this setting, regardless of noise level, DNN is able to classify some of them correctly, or at least as fundamentalists. Although the latter latter case is strongly affected by the noise level.
The chartists are also strongly affected by the features selection.
As expected, including features related to directed trends makes it relatively easy to classify them using both SVM and DNN.
Without these features classification performance drops, although chartists never really get mixed with other groups, not fundamentalists, nor noise traders.
An interesting observation here is that, for some chartists, increasing noise actually increases the F1-score for both SVM and DNN.
This may be an effect related to the fact that while classifying agents as chartists is easy, picking the correct specific sub-class is more difficult.
Such interpretation is in line with the effect being observed only for chartists with shorter horizons, as they tend to be more noisy on their own.
Finally, noise traders are well separated from the rest, but it should be noted that they are absent in the highest noise scenario.

\begin{table}
\begin{center}
\begin{minipage}{80mm}
\tbl{Classification Report for DNN architecture and 18 features.}
{\begin{tabular}{@{}c c c c c c c c c}\toprule
 & \multicolumn{4}{c}{\textbf{no-noise}} & \multicolumn{4}{c}{\textbf{66.6\% noise}} \\
\colrule
\textbf{Class} & \textbf{Precision} & \textbf{Recall} & \textbf{F1-score} & \textbf{Support} & \textbf{Precision} & \textbf{Recall} & \textbf{F1-score} & \textbf{Support} \\
\colrule
market maker(1) & 1.00 & 1.00 & 1.00 & 160 & 1.00 & 1.00 & 1.00 & 160 \\
market maker(2) & 1.00 & 1.00 & 1.00 & 160 & 1.00 & 1.00 & 1.00 & 160 \\
market maker(3) & 1.00 & 1.00 & 1.00 & 160 & 1.00 & 1.00 & 1.00 & 160 \\
market taker(1) & 0.99 & 0.99 & 0.99 & 80 & 0.90 & 0.91 & 0.91 & 80 \\
market taker(2) & 1.00 & 0.99 & 0.99 & 80 & 1.00 & 0.99 & 0.99 & 80 \\
market taker(3) & 0.98 & 1.00 & 0.99 & 80 & 0.90 & 0.91 & 0.91 & 80 \\
fundamentalist(1) & 0.49 & 0.64 & 0.55 & 80 & 0.48 & 0.53 & 0.50 & 80 \\
fundamentalist(2) & 0.74 & 0.79 & 0.76 & 80 & 0.32 & 0.14 & 0.19 & 80 \\
fundamentalist(3) & 0.82 & 0.78 & 0.79 & 80 & 0.13 & 0.03 & 0.04 & 80 \\
fundamentalist(4) & 0.59 & 0.44 & 0.50 & 80 & 0.44 & 0.35 & 0.39 & 80 \\
chartist(1) & 1.00 & 1.00 & 1.00 & 800 & 0.98 & 0.96 & 0.97 & 800 \\
chartist(2) & 1.00 & 1.00 & 1.00 & 800 & 0.83 & 0.95 & 0.89 & 800 \\
chartist(3) & 1.00 & 1.00 & 1.00 & 800 & 0.96 & 0.96 & 0.96 & 800 \\
chartist(4) & 1.00 & 1.00 & 1.00 & 800 & 0.89 & 0.92 & 0.90 & 800 \\
noise trader(1) & 1.00 & 1.00 & 1.00 & 8480 & N/A & N/A & N/A & N/A \\
\textbf{Overall Accuracy} & \multicolumn{4}{c}{\textbf{0.99}} & \multicolumn{4}{c}{\textbf{0.90}} \\ 
\botrule
\end{tabular}}
\label{tab:dnn_report_18}
\end{minipage}
\end{center}
\end{table}

To summarize, the classification is possible but requires sophisticated or informed features like, for example, looking at price trends.
It does not come as a surprise that the most challenging class of agents is the one made of traders who use external information, which is represented in our simulation as a fundamental price.
However, even then improvement is possible if some knowledge about the mechanism is known, like our assumption about fundamentalists activities near the fundamental price jumps.
The latter may represent an important economical announcement, for which we do not know the outcome, but we may know the publication time.
The examination of noise shows that it plays a role, but not as important as the correct selection of features.
What is important is that we know that the classes are separable, but before we move to clustering, let us shed some more light on how they are separated.

\begin{table}
\begin{center}
\begin{minipage}{80mm}
\tbl{Classification Report for DNN architecture and 9 features.}
{\begin{tabular}{@{}c c c c c c c c c}\toprule
 & \multicolumn{4}{c}{\textbf{no-noise}} & \multicolumn{4}{c}{\textbf{66.6\% noise}} \\
\colrule
\textbf{Class} & \textbf{Precision} & \textbf{Recall} & \textbf{F1-score} & \textbf{Support} & \textbf{Precision} & \textbf{Recall} & \textbf{F1-score} & \textbf{Support} \\
\colrule
market maker(1) & 1.00 & 1.00 & 1.00 & 160 & 1.00 & 1.00 & 1.00 & 160 \\
market maker(2) & 1.00 & 0.99 & 1.00 & 160 & 1.00 & 0.99 & 1.00 & 160 \\
market maker(3) & 0.99 & 1.00 & 1.00 & 160 & 0.99 & 1.00 & 1.00 & 160 \\
market taker(1) & 1.00 & 0.99 & 0.99 & 80 & 0.96 & 0.93 & 0.94 & 80 \\
market taker(2) & 1.00 & 1.00 & 1.00 & 80 & 1.00 & 1.00 & 1.00 & 80 \\
market taker(3) & 0.99 & 1.00 & 0.99 & 80 & 0.93 & 0.96 & 0.94 & 80 \\
fundamentalist(1) & 0.46 & 0.20 & 0.28 & 80 & 0.40 & 0.03 & 0.05 & 80 \\
fundamentalist(2) & 0.35 & 0.07 & 0.12 & 80 & 0.00 & 0.00 & 0.00 & 80 \\
fundamentalist(3) & 0.67 & 0.07 & 0.13 & 80 & 0.00 & 0.00 & 0.00 & 80 \\
fundamentalist(4) & 0.45 & 0.17 & 0.25 & 80 & 0.00 & 0.00 & 0.00 & 80 \\
chartist(1) & 0.55 & 0.37 & 0.45 & 800 & 0.46 & 0.60 & 0.52 & 800 \\
chartist(2) & 0.62 & 0.70 & 0.66 & 800 & 0.46 & 0.59 & 0.52 & 800 \\
chartist(3) & 0.57 & 0.34 & 0.42 & 800 & 0.53 & 0.46 & 0.49 & 800 \\
chartist(4) & 0.67 & 0.70 & 0.68 & 800 & 0.51 & 0.49 & 0.50 & 800 \\
noise trader(1) & 0.92 & 0.99 & 0.95 & 8480 & N/A & N/A & N/A & N/A \\
\textbf{Overall Accuracy} & \multicolumn{4}{c}{\textbf{0.85}} & \multicolumn{4}{c}{\textbf{0.57}} \\ 
\botrule
\end{tabular}}
\label{tab:dnn_report_9}
\end{minipage}
\end{center}
\end{table}

\begin{figure}
\begin{center}
\includegraphics[width=0.8\textwidth]{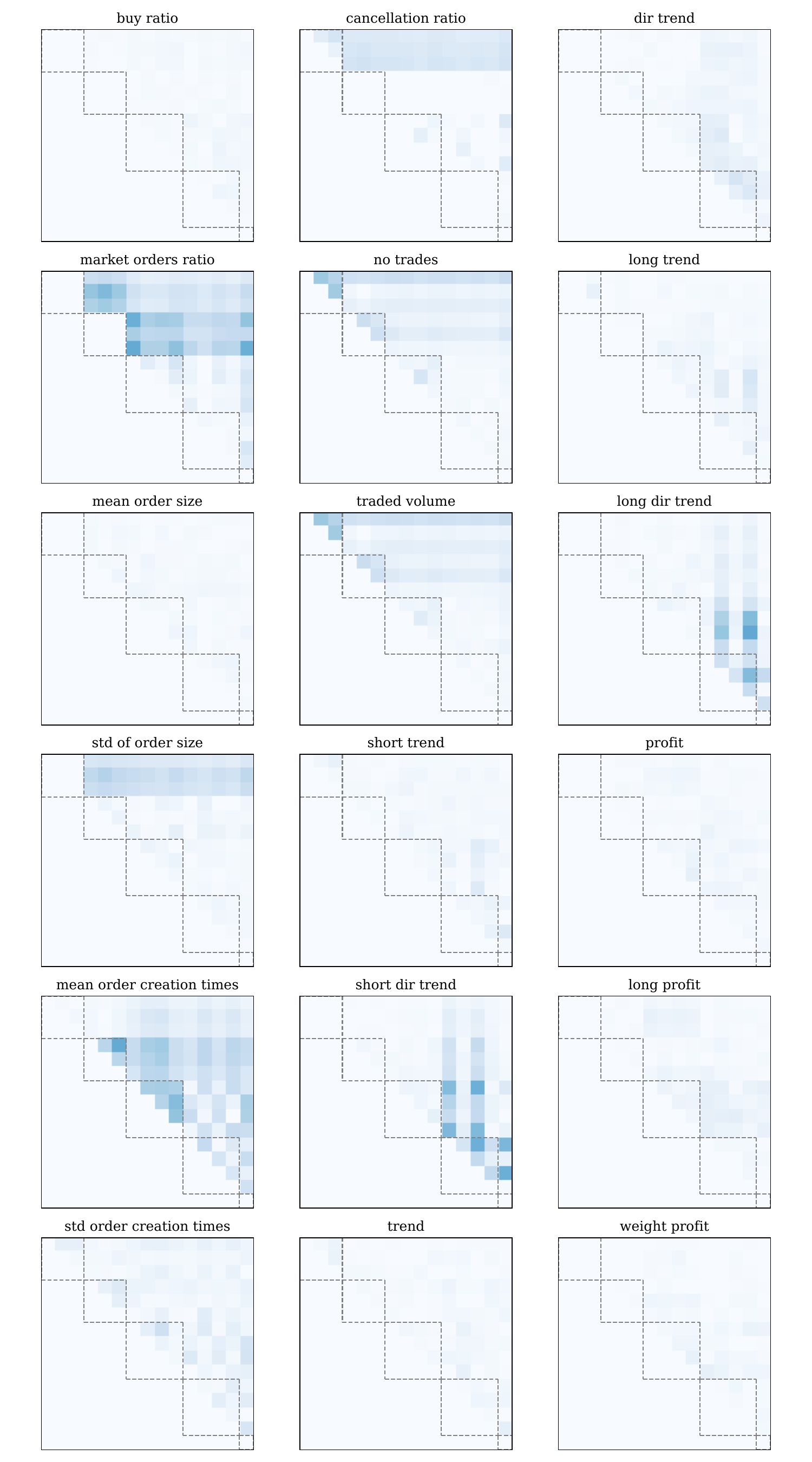}
\caption{Normalized weights for different features in a one-vs-one SVM scheme, in the setting of 18 features and no noise.
Dashed lines separate internal matrices for market makers (top left sub-matrix), market takers (sub-matrix second from the top), fundamentalists (sub-matrix third from the top) and chartists (down right sub-matrix).
\label{fig:svm_explain_18}}
\end{center}
\end{figure}

\begin{figure}
\begin{center}
\includegraphics[width=0.8\textwidth]{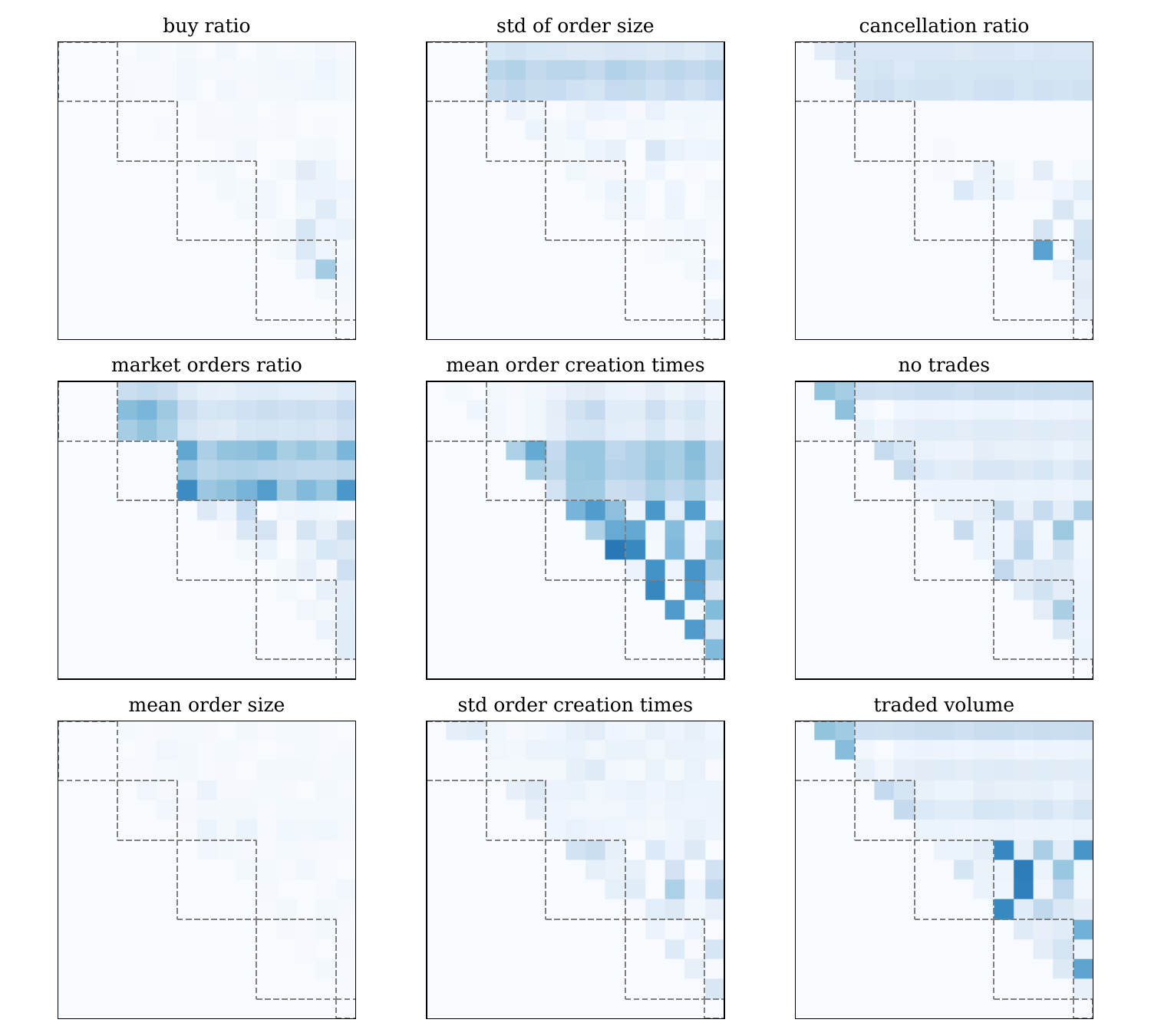}
\caption{Normalized weights for different features in a one-vs-one SVM scheme, in the setting of 9 features and no noise.
Dashed lines separate internal matrices for market makers (top left sub-matrix), market takers (sub-matrix second from the top), fundamentalists (sub-matrix third from the top) and chartists (down right sub-matrix).
\label{fig:svm_explain_9}}
\end{center}
\end{figure}

In order to understand how the supervised algorithms are capable of classifying the agents, we first look at the SVM weights for each pair of classes \citep{guyon2003introduction}.
Fig. \ref{fig:svm_explain_18} and Fig. \ref{fig:svm_explain_9} show for each feature a matrix of normalized absolute weights.
The normalization is done such that for each pair of classes the sum of weights over all features is equal to one.
All of these matrices are shown as upper triangular for readability, while understanding that the problem is symmetric.
This way a value in row $i$ and column $j$ should inform us on how important a certain feature is in distinguishing between class $i$ and $j$.
The order of rows and columns, as it was the case for confusion matrices, is the same as in Sec. \ref{sec:agents}.
Dashed squares were drawn to make it easier to spot dependencies between and inside different groups of classes.
We show only the noiseless case as adding noise does not change the matrices qualitatively and the conclusions stay the same.

Looking at both Fig. \ref{fig:svm_explain_18} and Fig. \ref{fig:svm_explain_9} we see that for SVM the cancellation ratio plays an important role in both distinguishing market makers from other classes as well as internally distinguishing between different types of market makers.
Standard deviation of orders' sizes helps in the former, but is not useful in the latter.
The ratio between the number of market and limit order plays a similar role as above, but for market takers.
The average order size is not utilized by SVM neither in the setting with 9 features nor for 18 features.
This does not come as a surprise since by construction agents post orders with size oscillating around five shares.
The ratio between buy and sell orders is weakly and without much success utilized by SVM in distinguishing between chartists, but no longer plays a role when 18 features are available.
The average order creation time has high SVM weights across different classes for 9 features setting, but it loses its importance when better features for distinguishing chartists and fundamentalists were available.
The standard deviation of the above order creation time has some influence on distinguishing chartists and fundamentalists in the 9 features setting, but becomes obsolete for the 18 features setting.
Both number of trades and the traded volume impact SVM in separating chartists and fundamentalists, as well as distinguishing different market makers.
While the latter holds for both settings with 9 and 18 features, the former is only observed for the smaller number of features available.
Similarity of their impact is a consequence of order sizes being in fact drew from a distribution centered around five shares.

When it comes to the additional features introduced in the manuscript, some of them significantly changed the distribution of weights and the quality of results, while others did not affect much the performance.
It comes as a surprise that trends, regardless of their length, are not particularly useful in this classification task.
This is most likely due to having the directed trend counterpart, which has more predictive power for chartists, but it is not as useful for other agents.
The latter not only dominated distinguishing between different types of chartists, but also between chartists and fundamentalists.
A closer look at Fig. \ref{fig:svm_explain_18} shows that among directed trend features with different lengths, the short and long ones are heavily utilized, but somehow the middle one is not as useful.
The reason is the same as before, other features seem to be already enough for the task.
The last three features, related to profit, were expected to help in classifying fundamentalists.
Their usefulness turned out to be rather limited, but they did help in distinguishing between different fundamentalists, especially the long profit.

\begin{figure}
\begin{center}
\includegraphics[width=\textwidth]{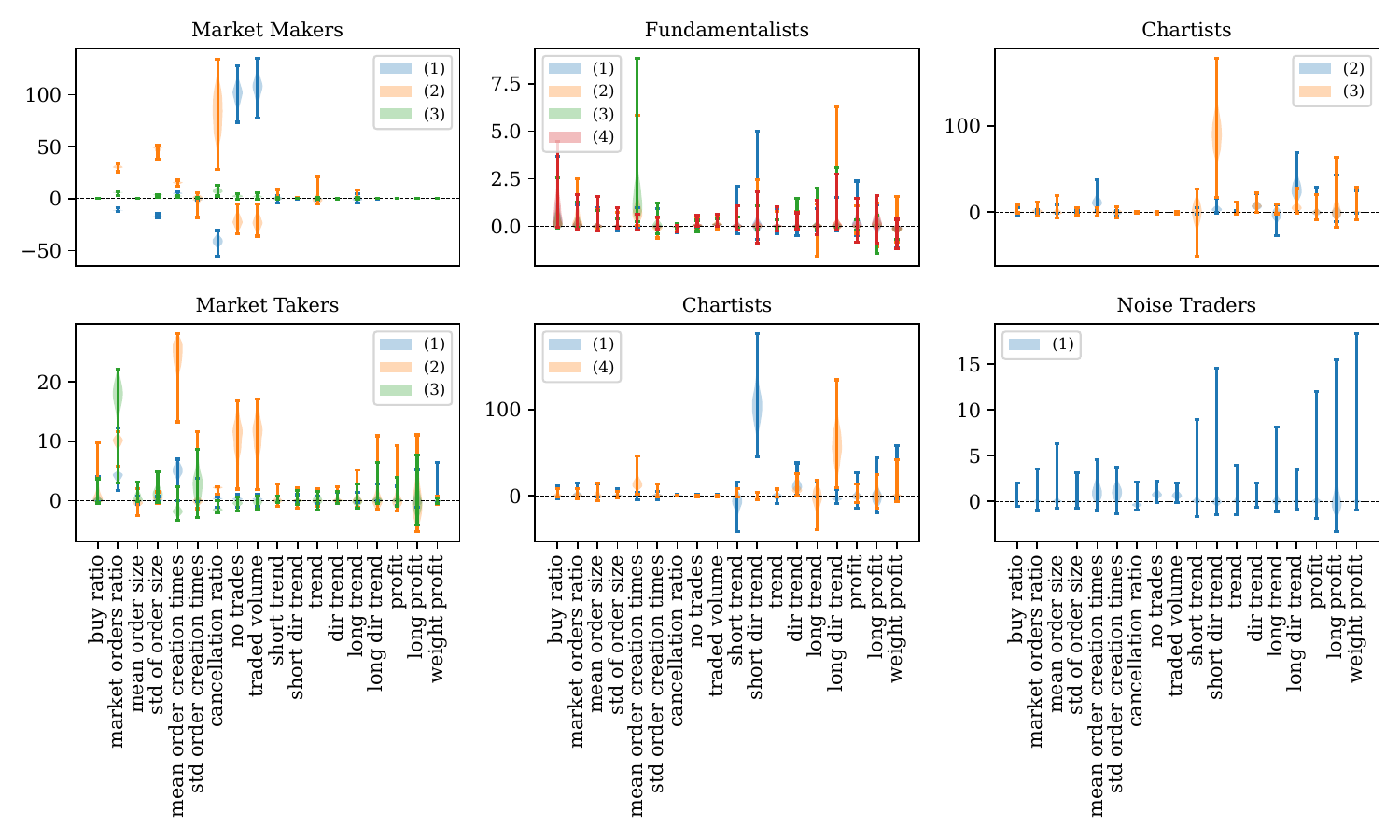}
\caption{Violin plots describing the distributions of \textit{relevance} \citep{montavon2019layer} of different features when predicting agents' classes using DNN.
Different colors depict different classes inside broader groups of agents.
Results obtained in the scenario with no additional noise and all 18 features.
\label{fig:dnn_explain_1_18}}
\end{center}
\end{figure}

Another attempt to understand which features play most important role in classifying agents is to use the layer-wise relevance propagation \citep{montavon2019layer}.
In this approach we measure how important certain features are in recognizing individual classes.
For market makers the most utilized features are the number of trades, the traded volume, and the cancellation ratio.
Market orders ratio and the deviation of order size also play a role, but their importance depends on the particular market maker's class.
As we introduce noise, the number of trades and the traded volume become more important in the classification task.
Since these are all basic features, the situation is almost the same for both 9 and 18 features.
In all scenarios DNN is very successful in classifying market makers.

\begin{figure}
\begin{center}
\includegraphics[width=\textwidth]{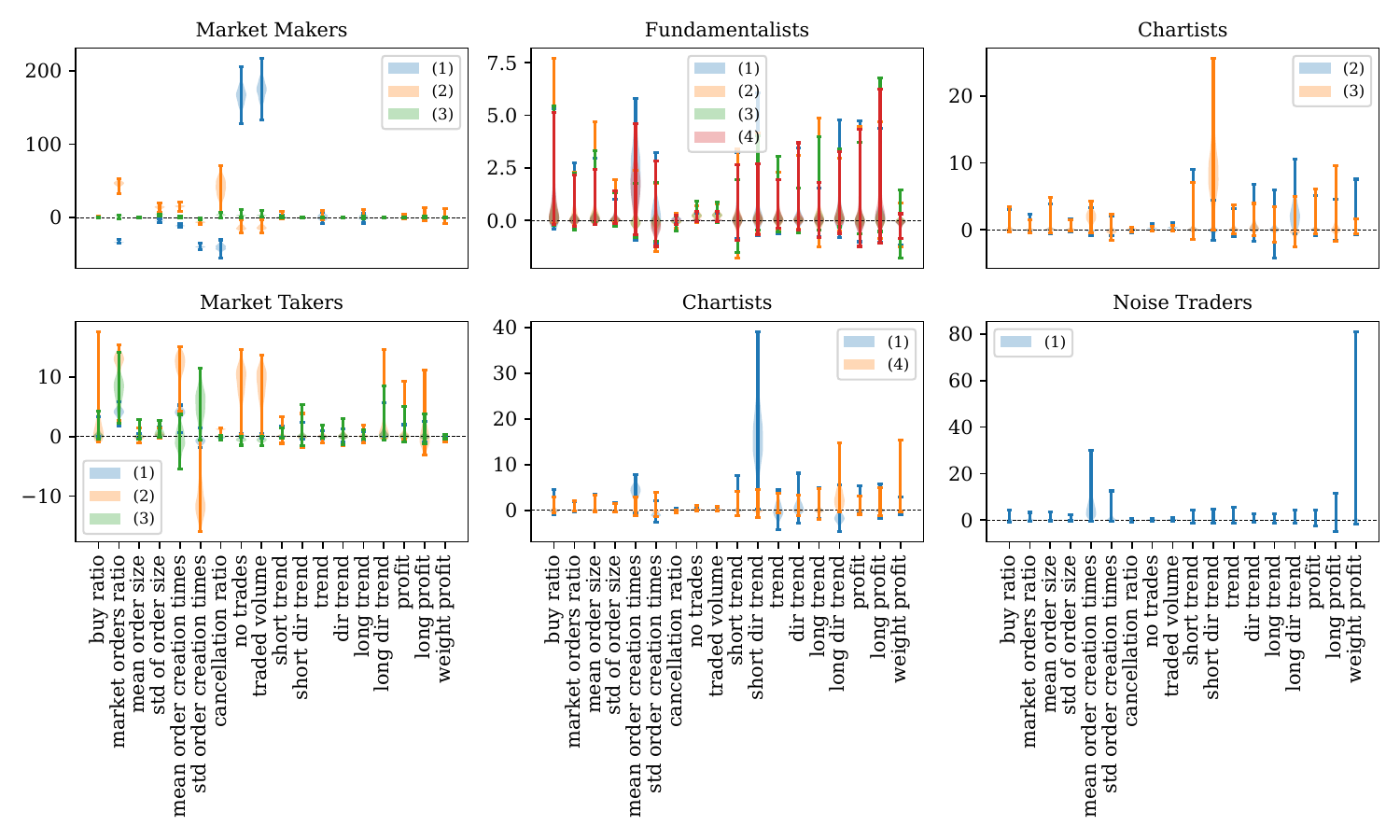}
\caption{Violin plots describing the distributions of \textit{relevance} \citep{montavon2019layer} of different features when predicting agents' classes using DNN.
Different colors depict different classes inside broader groups of agents.
Results obtained in the scenario with 50\% additional noise and all 18 features.
\label{fig:dnn_explain_2_18}}
\end{center}
\end{figure}

The most useful feature across different market takers is the ratio of market orders.
This does not come as a surprise, since market makers only use market orders, unless additional noise is added.
However, even in the latter case, this feature is heavily utilized by DNN.
Depending on the actual parametrization, average waiting time between orders can also be a strong indicator of being a market taker.
Similarly to the case of market makers, the number of trades and the traded volume are also seen as predictive features for both 9 and 18 features settings.
Interestingly, profit is also considered in classification, most likely due to the long time effects of their large orders.
Through a similar mechanism, long directional trend appears to be utilized as well.
All the mentioned features seem to have significant predictive power, as DNN is able to classify market takers quite easily across different settings.

\begin{figure}
\begin{center}
\includegraphics[width=\textwidth]{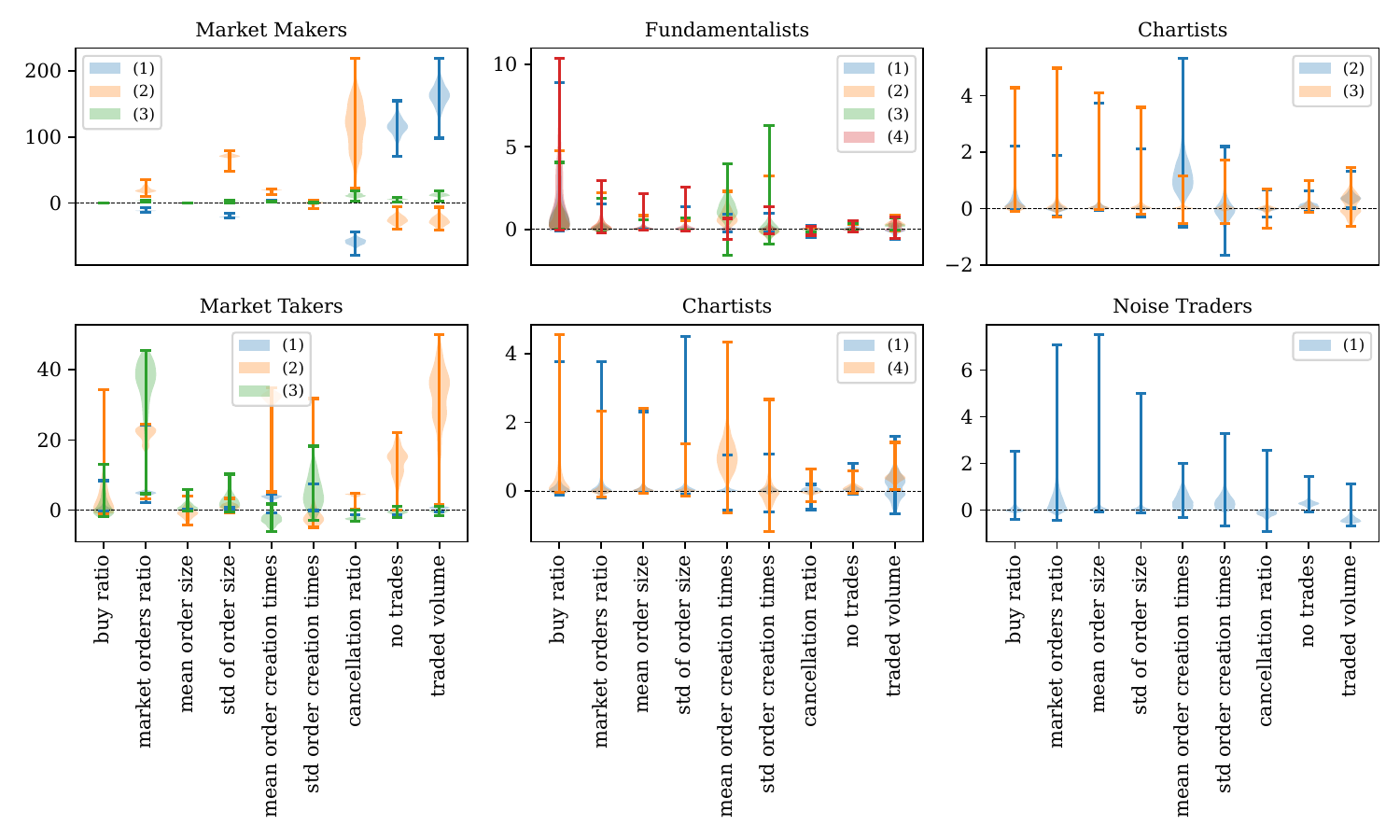}
\caption{Violin plots describing the distributions of \textit{relevance} \citep{montavon2019layer} of different features when predicting agents' classes using DNN.
Different colors depict different classes inside broader groups of agents.
Results obtained in the scenario with no additional noise and only 9 features.
\label{fig:dnn_explain_1_9}}
\end{center}
\end{figure}

As expected from the group which was most difficult to classify, the distribution of relevance for most features is centered around zero.
Only exception is the average orders' waiting time, which is most likely related to fundamentalists being active mostly around fundamental events represented by the jumps in the fundamental price.
The latter does not help in classification, however, which is confirmed by extremely low quality of the results with only 9 features.
In the more successful case of 18 features, the long directed trend appears to be used to classify fundamentalists.
This is kind of a self-fulfilling prophecy.
In other words, fundamentalists tend to push the price to the fundamental price, which is also a profitable direction for their orders.
While other agents, like chartists, have different views on the future price and so their activities average out in the macro scale, the fundamentalists are persistent as a group.
The analysis of relevance confirms the findings from the SVM weights, which is that the profit is not as useful in classifying fundamentalists as hoped.

As expected, directional trend plays a major role in classifying chartists.
Whether it is the short or the long version of trend that is utilized more, depends mostly on the chartist's parametrization.
When only 9 features are available, the average orders' waiting times are utilized by DNN, but it is not very successful in this scenario.

If we look closely at the noise traders, we notice that for them most of the features' relevance has a distribution which spans on both positive and negative sides.
This is partly expected, as actions of these agents are very random in their nature.
The most helpful features in this case, at least according to DNN, are related to orders' waiting times.
As seen for both 9 and 18 features scenarios, this is enough to classify noise traders with high accuracy.

\subsection{Clustering}

For clustering, we combine the training and validation sets, which together correspond to 70\% of the whole dataset, and use it for determining the clusters of agents.
The remaining 30\% is reserved for testing the performance of the clustering result, as it was done in the previous section.
The clustering hyper-parameters, including the linkage method and the number of clusters, are optimized using the noise-free dataset with all 18 features.
The same parameters are then applied to the datasets with noise (both with 9 and 18 features) and no noise scenario with 9 features.
We do not report the 50\% noise scenario as it provides similar observations as the other scenarios. 

A crucial aspect of hierarchical clustering is the linkage function used to calculate the distance between clusters and update the proximity matrix.
We conducted hierarchical clustering using Euclidean distance with four linkage methods: ward’s, complete, average, and centroid.
We analyzed the dendrograms and selected ward's method as it produces balanced and symmetric clusters with relatively uniform heights at the lower levels, ensuring compact and well-separated groups.
To further assess the validity of the clustering solution using ward's method, we compute the correlation between the cophenetic distances and the original distance matrix\footnote{Cophenetic distance is the height at which two observations are first joined together in a hierarchical clustering dendrogram, representing their similarity. A high correlation coefficient indicates that the clustering structure accurately reflects the pairwise distances of the original data. Generally, a correlation coefficient closer to 1 suggests a more valid clustering.}.
In our empirical analysis, we measured a 70\% correlation, indicating a fair agreement between the clustering structure and the original data distances.

\begin{figure}
\begin{center}
\begin{minipage}{\textwidth}
\subfigure[WCSS]{
\resizebox*{0.5\textwidth}{!}{\includegraphics{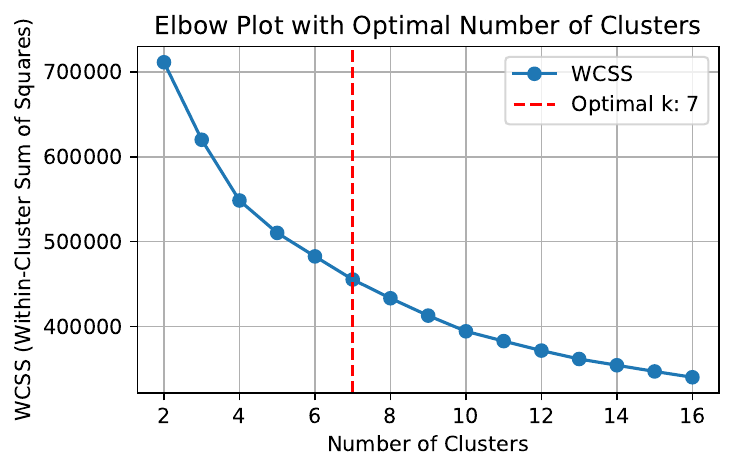}}\label{fig:wcss}}
\subfigure[Silhouette score]{
\resizebox*{0.5\textwidth}{!}{\includegraphics{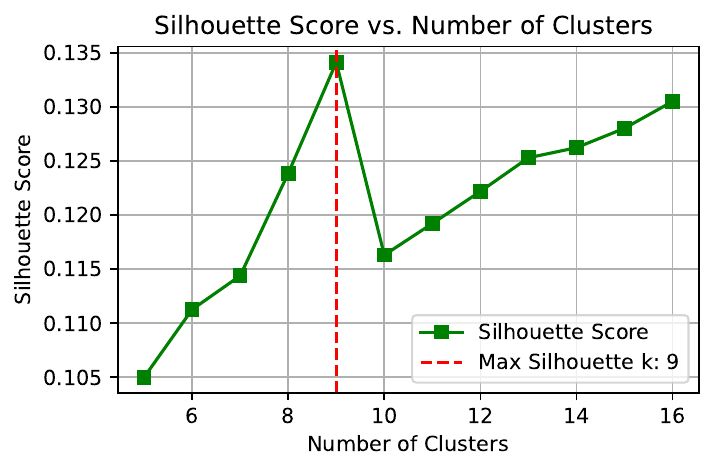}}\label{fig:sil}}
\caption{The Elbow Plot (left) illustrates the Within-Cluster Sum of Squares (WCSS) across different cluster numbers, with the optimal number of clusters determined using the Knee Locator method. The Silhouette Score Plot (right) evaluates cluster cohesion and separation, highlighting the number of clusters that maximizes the silhouette coefficient.).
\label{fig:elbow_plot}}
\end{minipage}
\end{center}
\end{figure}

To determine the optimal number of clusters, we evaluated the Silhouette coefficient and Within-Cluster Sum of Squares (WCSS).
Our empirical results show that the Silhouette coefficient peaked at 9 clusters ($S=0.134$) whereas the WCSS curve revealed an elbow point at $k=7$, where the rate of WCSS reduction slowed significantly (Fig. \ref{fig:elbow_plot}).
Taking these metrics into account, we identified \textbf{7} and \textbf{9} as the most suitable choices for the number of clusters.

To have a fair comparison with the supervised methods, we define the cluster composition.
Specifically, after clustering the training data, each cluster was assigned to the class that comprised the majority fraction within that cluster, aligning clusters and agent types.
In this way, we can have multiple classes of agents within one cluster.
However, using this approach, we may end up with empty clusters, i.e., some clusters may not have a clear dominant class.
To address this, we check the number of agents of each type within the empty cluster and the agent type with the highest count in that cluster is assigned to it.

Next, test samples were assigned to clusters by computing their minimum distance to the centroids obtained from the training data, ensuring consistency in cluster assignments across both datasets.
We then evaluated clustering accuracy on the test set, as summarized in Tables \ref{table:combined_results1}-\ref{table:combined_results4} for $k=9$ in both noise and no noise cases.{\footnote{Clustering accuracy for $k=7$ is not reported to avoid an overly lengthy paper.}
To compute precision, recall, and F1-score, we merge clusters corresponding to the same agent type.
Additionally, we report clustering accuracy for cases where the exact number of agent types is known, with $k=15$ in the noise-free scenario and $k=14$ in the presence of noise (see Tables \ref{table:combined_results1}-\ref{table:combined_results4}).

Firstly, we compare the performance of clustering with 18 features and 9 features, respectively, for the no noise case.
Clustering results for $k=9$ with 18 features show that market makers (1) were perfectly identified and assigned to Cluster 1 (F1 = 0.99), while market makers (2-3) were consistently grouped together, achieving an F1-score of 0.99.
This suggests that market makers exhibit sufficiently distinct behaviors, enabling the clustering method to capture their structure reliably.
However, the fact that market makers (2,3) were not assigned to separate clusters highlights the limitations of clustering in distinguishing between different subgroups of market makers.
In contrast, classification methods demonstrated significantly higher accuracy in identifying each subgroup.

Chartists, despite being split across multiple clusters, exhibited strong clustering performance, with F1-scores ranging from 0.93 to 0.94, confirming that their behavioral patterns were well captured when using the additional features introduced in this paper.
In contrast, fundamentalists (1-4) were grouped into a single cluster (Cluster 8, F1 = 0.47), indicating that clustering is not effective for distinguishing among these agent types.
Market takers (1-3) and noise traders were merged into Cluster 9 (F1 = 0.97), suggesting that their behaviors were too similar for clustering to differentiate them effectively.

For $k=9$ with 9 features, clustering accuracy dropped significantly from 94.39\% to 74.43\%.
This result clearly highlights the value and relevance of the additional features introduced in this paper.
While market makers remained well-separated (clusters 1 and 2, F1 $\approx$ 1.00), chartists and fundamentalists became increasingly mixed.
Chartist (2) and fundamentalists (2-3) were grouped together in Cluster 3 (F1 = 0.68), suggesting that the standard feature set lacks the ability to differentiate between them.
Similarly, chartist (4) was assigned to cluster 4 (F1 = 0.69), while fundamentalists (1,4) and chartist (1) were merged into cluster 6 (F1 = 0.24), confirming that the standard 9 features do not provide sufficient information to separate these agent types effectively.
This stands in sharp contrast to the 18-feature case, where chartists were identified with very high accuracy, demonstrating the necessity of incorporating the additional features.
Furthermore, noise traders and chartist (3) were now mixed within Cluster 7 (F1 = 0.84), indicating increased misclassification when using fewer features.

\begin{figure}[ht!]
\begin{center}
\begin{minipage}{\textwidth}
\subfigure[Scenario with 18 features and no noise with $k=9$. Overall accuracy is 95\%]{
\resizebox*{0.32\textwidth}{!}{\includegraphics{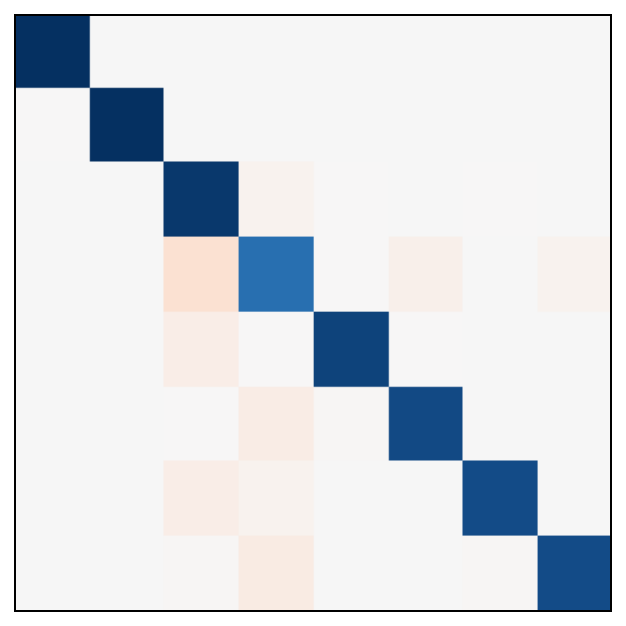}}\label{fig:c_1_18}}
\quad
\hspace{3.5cm} 
\subfigure[Scenario with 18 features and no noise with $k=15$. Overall accuracy is 75\%]{
\resizebox*{0.38\textwidth}{!}{\includegraphics{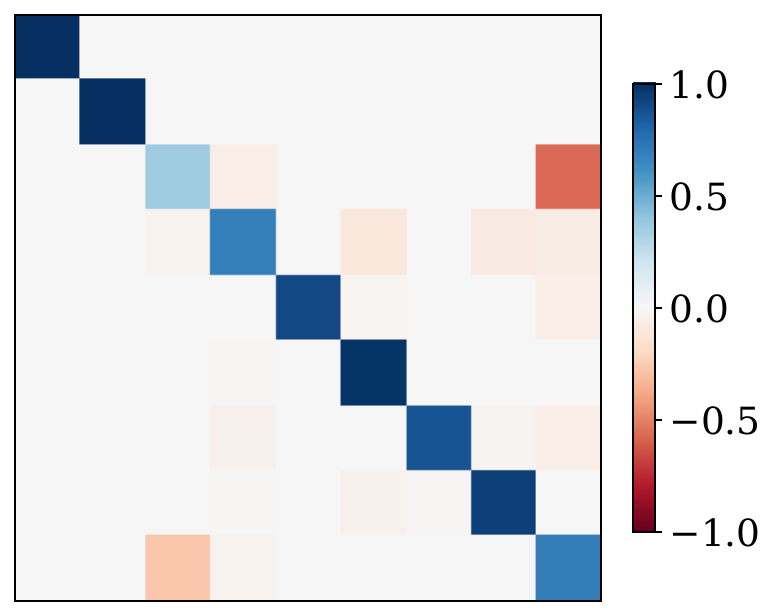}}\label{fig:c_2_18}}
\caption{Confusion matrix for clustering trading investors using Hierarchical clustering method with 18 features and no noise.
The order of clusters is the same as in Table \ref{table:combined_results1}.
\label{fig:hc1_18_cm}}
\end{minipage}
\end{center}
\end{figure}

\begin{table}[ht!]
\begin{center}
\begin{minipage}{80mm}
\tbl{Clustering report for no noise case (18 features) with Agent Type to Assigned Cluster Mapping}
{\begin{tabular}{@{}c c S[table-format=1.4] S[table-format=1.4] S[table-format=1.4] S[table-format=3.0]}\toprule
    
        \multicolumn{6}{c}{\textbf{Panel A: Number of Clusters $k=9$}} \\ \hline
        \textbf{Agent Type} & \textbf{Assigned Cluster} & \textbf{Precision} & \textbf{Recall} & \textbf{F1-Score} & \textbf{Support} \\ 
        \colrule
        market maker (1) & 1 & 0.9877 & 1.0000 & 0.9938 & 160 \\
        market maker (2-3) & 2 & 1.0000 & 0.9938 & 0.9969 & 320 \\
         noise trader (1), market taker (1-3) & 9 & 0.9801 & 0.9618 & 0.9709 & 8720 \\
        noise trader & 7 & & & & \\
        fundamentalist (1-4) & 8 & 0.3443 & 0.7531 & 0.4725 & 320 \\ 
       
        chartist (1) & 3 & 0.9623 & 0.9263 & 0.9439 & 800 \\
        chartist (2) & 4 & 0.9757 & 0.9050 & 0.9390 & 800 \\
        chartist (3) & 5 & 0.9677 & 0.8975 & 0.9313 & 800 \\
        chartist (4) & 6 & 0.9849 & 0.8975 & 0.9392 & 800 \\
        
        \textbf{Overall Accuracy} & \multicolumn{5}{c}{\textbf{0.9439}} \\ 
       \colrule
       
 \multicolumn{6}{c}{\textbf{Panel B: Number of Clusters $k=15$}} \\ \hline
        \textbf{Agent Type} & \textbf{Assigned Cluster} & \textbf{Precision} & \textbf{Recall} & \textbf{F1-Score} & \textbf{Support} \\ 
 \colrule
   
        market maker (1) & 1 & 0.9877 & 1.0000 & 0.9938 & 160 \\
        market maker (2-3) & 2 & 1.0000 & 0.9938 & 0.9969 & 320 \\
         market taker (1-3) & 14 & 0.0354 & 0.3583 & 0.0644 & 240 \\
 fundamentalist (1-4) & 13 & 0.3838 & 0.6813 & 0.4910 & 320 \\
       
        chartist (1) & 3 & 0.9770 & 0.9038 & 0.9390 & 800 \\
        chartist (2) & 4, 5, 6, 12 & 0.8992 & 0.9813 & 0.9384 & 800 \\
        chartist (3) & 7 & 0.9774 & 0.8650 & 0.9178 & 800 \\
        chartist (4) & 8, 9, 10 & 0.9328 & 0.9363 & 0.9345 & 800 \\
            
        noise trader (1) & 11, 15 & 0.9562 & 0.6900 & 0.8016 & 8480 \\
      
        \textbf{Overall Accuracy} & \multicolumn{5}{c}{\textbf{0.7533}} \\ 
      
  \botrule
\end{tabular}}
\label{table:combined_results1}
\end{minipage}
\end{center}
\end{table}

\begin{figure}[ht!]
\begin{center}
\begin{minipage}{\textwidth}
\subfigure[Scenario with 18 features and 66.6\% noise with $k=9$. Overall accuracy is 63\%]{
\resizebox*{0.32\textwidth}{!}{\includegraphics{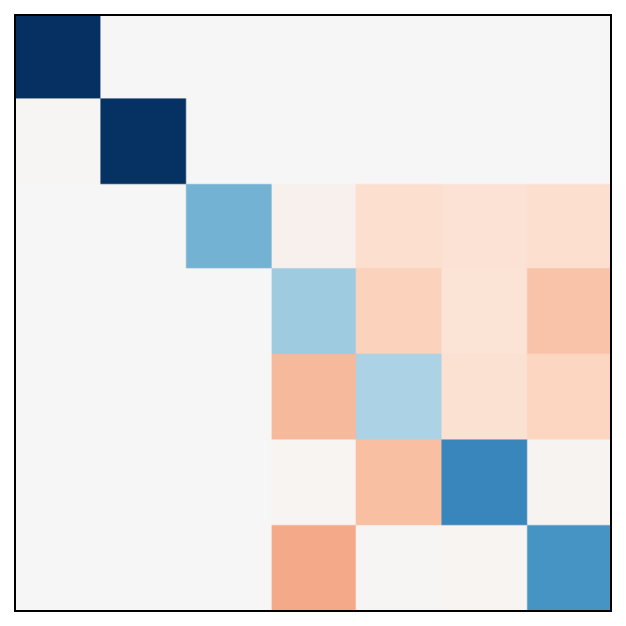}}\label{fig:cnoise_1_18}}
\quad
\hspace{3.5cm} 
\subfigure[Scenario with 18 features and 66.6\% noise with $k=14$. Overall accuracy is 91\%]{
\resizebox*{0.38\textwidth}{!}{\includegraphics{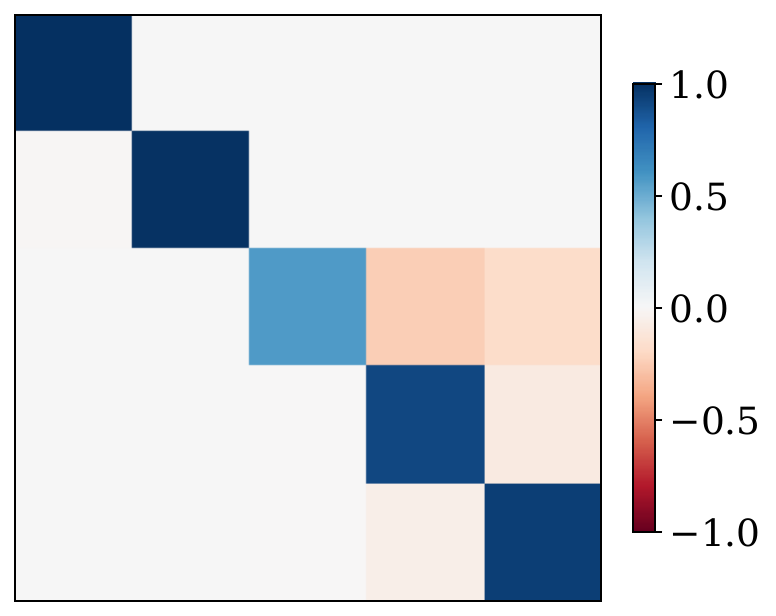}}\label{fig:cnoise_2_18}}
\caption{Confusion matrix for clustering trading investors using Hierarchical clustering method with 18 features and noise.
The order of clusters is the same as in Table \ref{table:combined_results2}.
\label{fig:hc2_18_cm}}
\end{minipage}
\end{center}
\end{figure}

\begin{table}
\begin{center}
\begin{minipage}{80mm}
\tbl{Clustering report for noise case (18 features) with Agent Type to Assigned Cluster Mapping}
{\begin{tabular}{@{}c c S[table-format=1.4] S[table-format=1.4] S[table-format=1.4] S[table-format=3.0]}\toprule
    
        \multicolumn{6}{c}{\textbf{Panel A: Number of Clusters $k=9$}} \\ \hline
        \textbf{Agent Type} & \textbf{Assigned Cluster} & \textbf{Precision} & \textbf{Recall} & \textbf{F1-Score} & \textbf{Support} \\ 
        \colrule       
        market maker (1) & 1 & 0.9756 & 1.0000 & 0.9877 & 160 \\
        market maker (2-3) & 2 & 1.0000 & 0.9844 & 0.9921 & 320 \\
          market taker (1-3) & 4 & 0.9661 & 0.4750 & 0.6369 & 240 \\
          fundamentalist (1-3) & 5 & 0.1108 & 0.3542 & 0.1688 & 240 \\
        fundamentalist (4) & 6 & 0.0403 & 0.3125 & 0.0714 & 80 \\
  
        chartist (1-2) & 3 & 0.9092 & 0.6506 & 0.7585 & 1600 \\
        chartist (2) & 8 & & & & \\
        chartist (3-4) & 7 & 0.8461 & 0.5875 & 0.6935 & 1600 \\
        chartist (4)& 9 & & & & \\

        \textbf{Overall Accuracy} & \multicolumn{5}{c}{\textbf{0.6321}} \\ 

        \colrule
        \multicolumn{6}{c}{\textbf{Panel B: Number of Clusters $k=14$}} \\ \hline
        \textbf{Agent Type} & \textbf{Assigned Cluster} & \textbf{Precision} & \textbf{Recall} & \textbf{F1-Score} & \textbf{Support} \\
        \colrule
        market maker (1) & 1 & 0.9756 & 1.0000 & 0.9877 & 160 \\
        market maker (2-3) & 3 & 1.0000 & 0.9875 & 0.9937 & 320 \\
        market maker (3) & 2 & & & & \\
           market taker (1-3) & 5& 0.9643 & 0.5625 & 0.7105 & 240 \\
        market taker (2) &  8 &&&&\\
  fundamentalist (1-4), chartist (3, 4) & 6 & 0.9175 & 0.9089 & 0.9131 & 1920 \\
       chartist (3) &  10, 11, 14 & & & & \\
        chartist (4) &  13 & & & & \\

        chartist (1, 2) & 4& 0.8737 & 0.9381 & 0.9048 & 1600 \\
        chartist (2) & 7, 12 & & & & \\
        chartist (1) & 9 & & & & \\
       
        \textbf{Overall Accuracy} & \multicolumn{5}{c}{\textbf{0.9097}} \\ 
        \botrule
\end{tabular}}
\label{table:combined_results2}
\end{minipage}
\end{center}
\end{table}

The results for $k=15$ with 18 features show that increasing the number of clusters had mixed effects.
For 18 features, finer clustering improved chartist (2) and (4) separation, with chartist (2) achieving high recall (0.98) across clusters 4, 5, 6, and 12.
Market takers (1-3) were assigned to cluster 14 with an extremely low F1-score (0.06), showing that increasing $k$ led to greater dispersion rather than improved classification.
Fundamentalists (1-4) remained difficult to separate (F1 = 0.49), and noise traders were split across clusters 11 and 15 (F1 = 0.80), reducing their distinction.
As a result, overall accuracy dropped from 94.39\% ($k=9$) to 75.33\% ($k=15$), confirming that increasing $k$ did not necessarily improve clustering performance.
For 9 features, an unexpected advantage emerged: market makers (2-3), which were indistinguishable in the 18-feature case, were now correctly assigned to distinct clusters (F1 = 1.00).
However, this benefit came at a cost— chartist (2) and fundamentalist (2) were merged (F1 = 0.55), and fundamentalist (3) suffered even lower classification performance (F1 = 0.19), suggesting that fewer features made it harder to distinguish these groups.
Noise traders remained relatively distinct (F1 = 0.88) but showed weaker separation from chartists than in the case of 18 features.

\begin{figure}[ht!]
\begin{center}
\begin{minipage}{\textwidth}
\subfigure[Scenario with 9 features and no noise with $k=9$. Overall accuracy is 74\%]{
\resizebox*{0.31\textwidth}{!}{\includegraphics{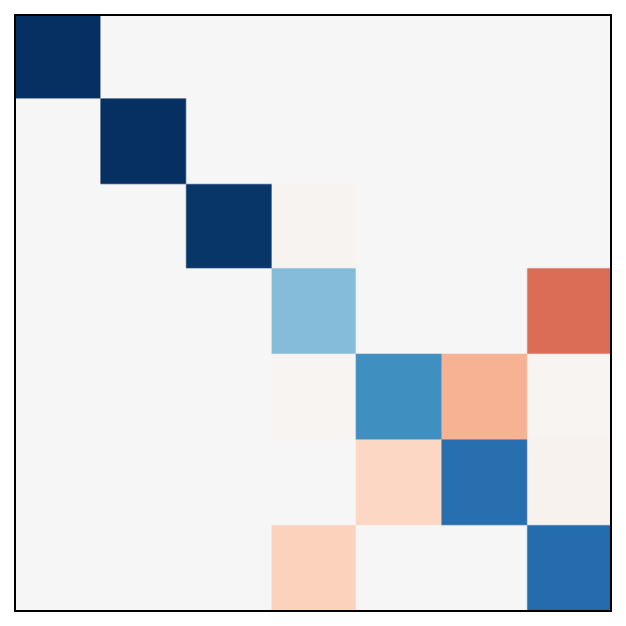}}\label{fig:c_1_9}}
\quad
\hspace{3.5cm} 
\subfigure[Scenario with 9 features and no noise with $k=15$. Overall accuracy is 76\%]{
\resizebox*{0.38\textwidth}{!}{\includegraphics{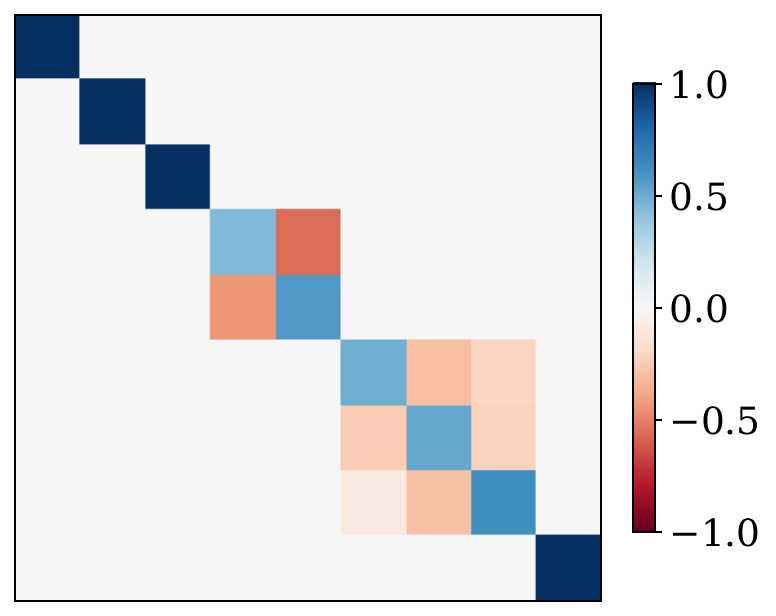}}\label{fig:c_2_9}}
\caption{Confusion matrix for clustering trading investors using Hierarchical clustering method with 9 features and no noise.
The order of clusters is the same as in Table \ref{table:combined_results3}.
\label{fig:hc3_18_cm}}
\end{minipage}
\end{center}
\end{figure}

\begin{table} [ht!]
\begin{center}
\begin{minipage}{80mm}
\tbl{Clustering report for  no noise case (9 features) with Agent Type to Assigned Cluster Mapping}
{\begin{tabular}{@{}c c S[table-format=1.4] S[table-format=1.4] S[table-format=1.4] S[table-format=3.0]}\toprule
        \multicolumn{6}{c}{\textbf{Panel A: Number of Clusters $k=9$}} \\ \hline
        \textbf{Agent Type} & \textbf{Assigned Cluster} & \textbf{Precision} & \textbf{Recall} & \textbf{F1-Score} & \textbf{Support} \\ 
        \colrule       

        market maker (1) & 1 & 0.9877 & 1.0000 & 0.9938 & 160 \\
        market maker (2-3) & 2 & 1.0000 & 0.9938 & 0.9969 & 320 \\
         market taker (1-3) & 9 & 1.0000 & 0.9708 & 0.9852 & 240 \\
      fundamentalist (1, 4), chartist (1) & 6 & 0.1626 & 0.4365 & 0.2369 & 960 \\
      
         fundamentalist (2-3), chartist (2) & 3 & 0.7768 & 0.6125 & 0.6849 & 960 \\
        chartist (4) & 4 & 0.6426 & 0.7550 & 0.6943 & 800 \\
        noise trader (1), chartist (3) & 7 & 0.9241 & 0.7700 & 0.8401 & 9280 \\
        noise trader (1) & 5, 8 & & & & \\
       
        \textbf{Overall Accuracy} & \multicolumn{5}{c}{\textbf{0.7443}} \\ 
        
        \colrule
        \multicolumn{6}{c}{\textbf{Panel B: Number of Clusters $k=15$}} \\ \hline
        \textbf{Agent Type} & \textbf{Assigned Cluster} & \textbf{Precision} & \textbf{Recall} & \textbf{F1-Score} & \textbf{Support} \\ 
        \colrule
        market maker (1) & 1 & 1.0000 & 1.0000 & 1.0000 & 160 \\
        market maker (3) & 2 & 1.0000 & 1.0000 & 1.0000 & 160 \\
        market maker (2) & 3 & 1.0000 & 1.0000 & 1.0000 & 160 \\
           market taker (1) & 14 & 0.3302 & 0.4375 & 0.3763 & 80 \\
        market taker (2-3) & 15 & 0.6691 & 0.5688 & 0.6149 & 160 \\
fundamentalist (3) & 5 & 0.1175 & 0.4875 & 0.1893 & 80 \\
  fundamentalist (1, 4), chartist (1) & 11 & 0.1513 & 0.2583 & 0.1908 & 960 \\
       
      fundamentalist (2), chartist (2) & 4 & 0.6227 & 0.5045 & 0.5574 & 880 \\
        chartist (2) & 7 & & & & \\
        chartist (4) & 6, 8 & 0.6907 & 0.6113 & 0.6485 & 800 \\
        noise trader (1), chartist (3) &  13 & 0.9170 & 0.8504 & 0.8825 & 9280 \\
        noise trader (1) &  9, 10, 13 & & & & \\

        \textbf{Overall Accuracy} & \multicolumn{5}{c}{\textbf{0.7640}} \\ 
       
        \botrule
\end{tabular}}
\label{table:combined_results3}
\end{minipage}
\end{center}
\end{table}

The introduction of noise significantly impacted clustering accuracy, with most agent types becoming harder to classify.
For $k=9$ with 18 features, accuracy dropped from 94.39\% to 63.21\%, as chartists mixed across clusters and fundamentalists became nearly indistinguishable (F1 = 0.17).
Market takers and noise traders were also harder to separate, reinforcing that clustering struggles to maintain meaningful structure under noise.
A similar but less severe decline occurred in the 9-feature case ($k=9$, 74.43\% to 62.00\%), suggesting that trend-based features (only in the case of 18 features) are more affected by noise, making agent behaviors less distinguishable.

\begin{figure}[ht!]
\begin{center}
\begin{minipage}{\textwidth}
\subfigure[Scenario with 9 features and 66.6\% noise with $k=9$. Overall accuracy is 62\%]{
\resizebox*{0.32\textwidth}{!}{\includegraphics{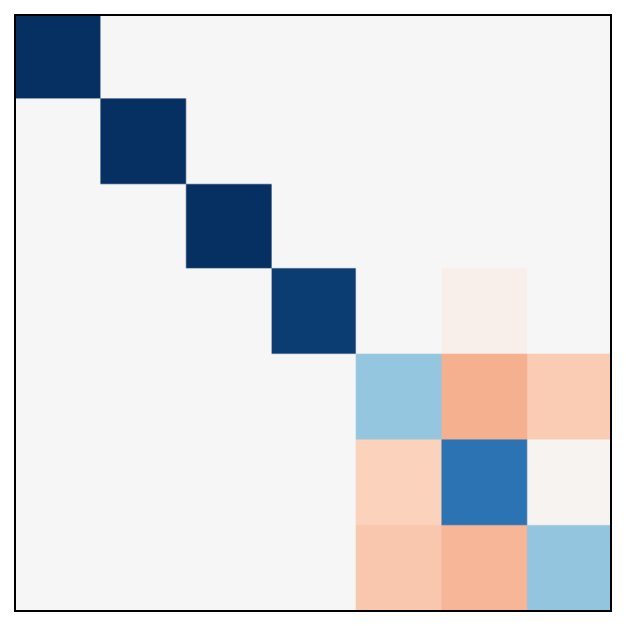}}\label{fig:cnoise_1_9}}
\quad
\hspace{3.5cm} 
\subfigure[Scenario with 9 features and 66.6\% noise with $k=14$. Overall accuracy is 53\%]{
\resizebox*{0.4\textwidth}{!}{\includegraphics{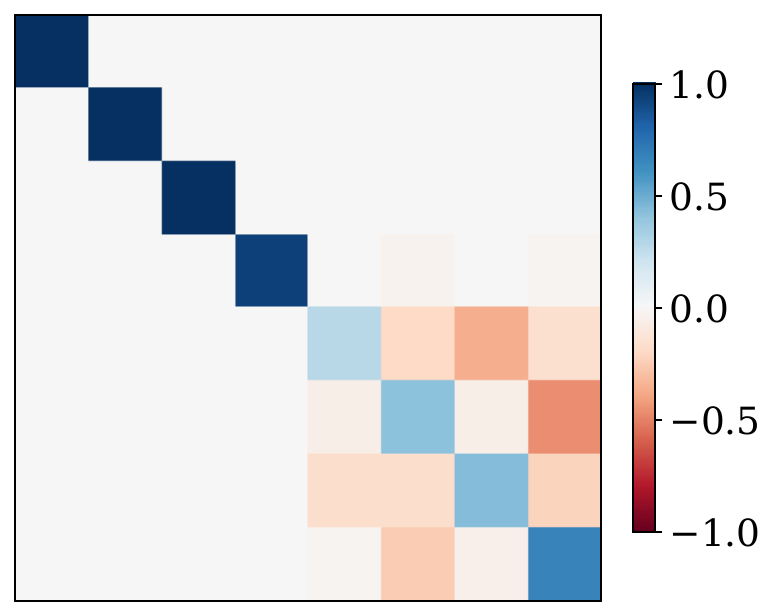}}\label{fig:cnoise_2_9}}
\caption{Confusion matrix for clustering trading investors using Hierarchical clustering method with 9 features and noise.
The order of clusters is the same as in Table \ref{table:combined_results4}.
\label{fig:hc4_18_cm}}
\end{minipage}
\end{center}
\end{figure}

\begin{table}[ht!]
\begin{center}
\begin{minipage}{80mm}
\tbl{Clustering report for noise case (9 features) with Agent Type to Assigned Cluster Mapping}
{\begin{tabular}{@{}c c S[table-format=1.4] S[table-format=1.4] S[table-format=1.4] S[table-format=3.0]}\toprule
    
        \multicolumn{6}{c}{\textbf{Panel A: Number of Clusters $k=9$}} \\ \hline
        \textbf{Agent Type} & \textbf{Assigned Cluster} & \textbf{Precision} & \textbf{Recall} & \textbf{F1-Score} & \textbf{Support} \\ 
        \colrule      
        market maker (1) & 7 & 1.0000 & 1.0000 & 1.0000 & 160 \\
        market maker (3) & 8 & 1.0000 & 1.0000 & 1.0000 & 160 \\
        market maker (2) & 9 & 1.0000 & 1.0000 & 1.0000 & 160 \\
     market taker (1-3) &  6 & 0.9827 & 0.9458 & 0.9639 & 240 \\
        market taker (2) & 5 & & & & \\
fundamentalist (1-4), chartist (2) & 4 & 0.4212 & 0.3839 & 0.4017 & 1120 \\
       
        chartist (1, 3) &  3 & 0.6328 & 0.7356 & 0.6803 & 1600 \\
        chartist (1) & 1 & & & & \\
        chartist (4) & 2 & 0.4861 & 0.3938 & 0.4351 & 800 \\
               
        \textbf{Overall Accuracy} & \multicolumn{5}{c}{\textbf{0.6200}} \\ 
      \colrule
        
  \multicolumn{6}{c}{\textbf{Panel B: Number of Clusters $k=14$}} \\ \hline
        \textbf{Agent Type} & \textbf{Assigned Cluster} & \textbf{Precision} & \textbf{Recall} & \textbf{F1-Score} & \textbf{Support} \\ 
        \colrule
        market maker (1) & 12 & 1.0000 & 1.0000 & 1.0000 & 160 \\
        market maker (3) & 13 & 1.0000 & 1.0000 & 1.0000 & 160 \\
        market maker (2) & 14 & 1.0000 & 1.0000 & 1.0000 & 160 \\
          market taker (1-3) & 11 & 0.9868 & 0.9333 & 0.9593 & 240 \\
        market taker (2) & 9,10 & & & & \\ 
       fundamentalist (2-3) & 8 & 0.1076 & 0.2750 & 0.1547 & 160 \\
       
     fundamentalist (1, 4), chartist (1)  &  7 & 0.4315 & 0.4104 & 0.4207 & 960 \\
        chartist (1) & 1 & & & & \\
        chartist (2, 4) & 2 & 0.8195 & 0.4313 & 0.5651 & 1600 \\
        chartist (2) & 6 & & & &\\
        chartist (3) & 3, 4, 5 & 0.3908 & 0.6688 & 0.4933 & 800 \\
        
        \textbf{Overall Accuracy} & \multicolumn{5}{c}{\textbf{0.5583}} \\ 
         
        \botrule
\end{tabular}}
\label{table:combined_results4}
\end{minipage}
\end{center}
\end{table}

For $k=15$ with 18 features, accuracy remained high (90.97\%), but this was due to more merging of similar agent types rather than improved separation.
Unlike in the no-noise case, where fragmentation increased with $k=15$, noise caused more agent groups to collapse into fewer clusters, artificially inflating accuracy.
In contrast, for $k=15$ with 9 features, accuracy fell sharply (76.40\% to 55.83\%), confirming that clustering in noisy environments tends to either merge multiple groups (as in 18 features) or create excessive fragmentation (as in 9 features).
Overall, these results highlight that clustering is not robust under noise, as it either loses distinctions between agent types or fails to maintain consistent separations. 

We would like to highlight the fact that, in our case, the clustering strategy had multiple agent types in the same cluster.
If we were to break these clusters into their constituent classes, the accuracy would decline significantly.
Clustering performance is already inferior to the classification performance and adopting a more rigid approach for assigning agents to clusters would further increase the gap.

\section{Discussion and Conclusions}

Before diving into discussion about the results, we need to comment on the relation between agents' parametrization and the effectiveness of using certain features in classifying agents.
It is clear that some features will not bring much information due to the specific parametrization of agents.
One example would be the mean order size, which is the same for all agents.
This is done on purpose as we did not want to make the task of classification or clustering too simple, when it was not necessary.
Some features, on the other hand, were strongly indicative of certain classes, because of the specifics of agents behavior.
This includes market order ratio, which made it easier to classify market makers and market takers.
This is partly changed through adding noise, but as we see in the results, classifying them is still remarkably successful.
In some cases, even if parameters partly control agent's activity patterns, it is still highly dependent on the market conditions, like for chartists.
Finally, in the case of fundamentalists, their behavior is conditioned on both market dynamics as well as external factors described by the fundamental price.

What the results show clearly is that classification of agents in our simulation setting is possible with high accuracy.
Some agents, like market makers and market takers, are relatively easy to distinguish from others due to their distinctive trading behavior.
Others, like chartists or to a certain extent fundamentalists, require more informed features.
The above statement is robust in terms of additional noise, which has a small, but visible effect almost entirely for fundamentalists.
It needs to be highlighter though, that fundamentalists model investors, which behavior is a function of external factors and as such are extremely difficult to classify using only trading data.
Important takeaway message, confirmed by both tested methods, SVM and DNN, is that there is a dependence, which we are able to utilize using a supervised setting.
This dependence allows us to classify agents accurately.
Now, the question is whether we can do the same in an unsupervised manner.

Comparing classification and clustering results is not an easy task.
For one, there is a set of information, which is easily available for supervised classification, and difficult to obtain for unsupervised clustering.
One good example is the number of classes, which is given in the first case, but unknown for clustering algorithms.
Moreover, even if one somehow decides on the number of clusters, they may not correspond directly to classes or even subsets of classes.
This is the case in our experiments were both clusters could contain number of classes as well as classes could be part of multiple clusters.
For our purpose, however, the big picture is clear despite the limitations we encountered.
Clustering methods are not capable of achieving the levels of accuracy, which we observe in the supervised scenario.
Even if overall accuracy is high, like in the case of 18 features and no noise, it is at the expense of merging multiple classes into a single cluster.
This way we end up clustering together agents as diverse as noise traders and market takers.
If we then forcefully try to divide classes into separate clusters, by setting the correct number of clusters, the results become much less accurate.
The only case where the results of clustering and classification are comparable is when we fix the correct number of clusters and use the noisy data.
This is simply a consequence of merging fundamentalists and chartists in the process.
For classification, they are the sole source of misclassification and if we merge them, the problem entirely disappears.

While working on the problem, we noticed that not only is clustering unable to fully reproduce the dependencies we find using supervised approach, it is also much more limited in terms of the size of the dataset.
In finance, where datasets are large and the number of investors can be orders of magnitude bigger than what we have in our simulations, this is a problem.
This highlights a need for clustering methods capable of dealing with much larger datasets.

In summary, we have shown that standard features may have difficulties in distinguishing investors with more sophisticated strategies.
Additionally, we suggested targeted features to deal with that problem, i.e. trend-based features allowed to distinguish chartists from other agents.
We also investigated investors with external information and used features based on partial information about external events, in order to improve the results.
Finally, on top of classification being systematically better than clustering, it is important to highlight that noise have stronger impact on the latter.

Our study is limited and does not take into account all elements of the market or realities of available data.
This includes modeling of the noise, which can be a consequence of data quality, as well as human behavior.
Moreover, in reality investors are not really grouped in separate classes with distinct trading behaviors.
They may for example take different roles at different times or shift from one strategy to another.
Even if their strategy is similar, the underlying data, often external, may be very different or may be interpreted differently.
Furthermore,there are endless possibilities when it comes to setting, methods and parametrization.
We could have more agents, divide them into more classes with different activity patterns, and use multiple methods for both supervised and unsupervised scenarios.
Nevertheless, the conclusion is clear and will most likely stay the same under different conditions and for other methods.
Actually, we would expect the results to deteriorate under more realistic conditions, making the message even stronger.

That said, there are multiple future directions, which we find interesting and hope to pursue in the future.
One is to extend the ABM further, especially in the direction of external factors, such as interaction between investors, which takes places outside of the market.
While we know that external factors make the classification difficult, external interactions can be sometimes reconstructed from the data \citep{baltakiene2021identification} and serve as an additional feature.
One direction, which could address some of the issues described in the manuscript, but would require more information, is to have a semi-supervised learning approach.
If, for example, we know that certain events are connected to official market makers, this could inform the whole procedure.
Finally, we cannot forget about the fast pace of development in ML these days, which may lead to developing methods capable of clustering the data in a more efficient and accurate way.
All these points confirm our belief that ABM is a great tool for validating new methods and approaches in finance, where experimental capabilities are still very limited.

\section*{Acknowledgements}
M. Wilinski acknowledges support from the European Union’s Horizon Europe programme under the Marie Skłodowska-Curie Actions (Grant Agreement No. 101066936,
Project: DataABM).
A. Iosifidis acknowledges funding from the Independent Research Fund Denmark project DeepFINA (grant ID 10.46540/3105-00031B).

\vspace{-0.5cm}

\bibliographystyle{rQUF}
\bibliography{refs}

\begin{thebibliography}{38}
\providecommand{\natexlab}[1]{#1}
\providecommand{\noopsort}[1]{}
\providecommand{\printfirst}[2]{#1}
\providecommand{\singleletter}[1]{#1}
\providecommand{\switchargs}[2]{#2#1}

\bibitem[\protect\citeauthoryear{Abhyankar {\itshape{et~al.}}}{1997}]{abhyankar1997bid}
Abhyankar, A., Ghosh, D., Levin, E. and Limmack, R., Bid-ask spreads, trading volume and volatility: Intra-day evidence from the London Stock Exchange. {\itshape Journal of Business Finance \& Accounting}, 1997, \textbf{24}, 343--362.

\bibitem[\protect\citeauthoryear{Baltakien{\.e} {\itshape{et~al.}}}{2021}]{baltakiene2021identification}
Baltakien{\.e}, M., Kanniainen, J. and Baltakys, K., Identification of information networks in stock markets. {\itshape Journal of Economic Dynamics and Control}, 2021, \textbf{131}, 104217.

\bibitem[\protect\citeauthoryear{Belcak {\itshape{et~al.}}}{2021}]{belcak2021fast}
Belcak, P., Calliess, J.P. and Zohren, S., Fast Agent-Based Simulation Framework with Applications to Reinforcement Learning and the Study of Trading Latency Effects. In {\itshape Proceedings of the }{\itshape International Workshop on Multi-Agent Systems and Agent-Based Simulation}, pp. 42--56, 2021.

\bibitem[\protect\citeauthoryear{Bouchaud {\itshape{et~al.}}}{2002}]{bouchaud2002statistical}
Bouchaud, J.P., M{\'e}zard, M. and Potters, M., Statistical properties of stock order books: empirical results and models. {\itshape Quantitative finance}, 2002, \textbf{2}, 251.

\bibitem[\protect\citeauthoryear{Brogaard {\itshape{et~al.}}}{2014}]{brogaard2014high}
Brogaard, J., Hendershott, T. and Riordan, R., High-Frequency Trading and Price Discovery. {\itshape Review of Financial Studies}, 2014.

\bibitem[\protect\citeauthoryear{Byrd {\itshape{et~al.}}}{2020}]{byrd2020abides}
Byrd, D., Hybinette, M. and Balch, T.H., ABIDES: Towards high-fidelity multi-agent market simulation. In {\itshape Proceedings of the }{\itshape Proceedings of the 2020 ACM SIGSIM Conference on Principles of Advanced Discrete Simulation}, pp. 11--22, 2020.

\bibitem[\protect\citeauthoryear{Cao {\itshape{et~al.}}}{2009}]{cao2009information}
Cao, C., Hansch, O. and Wang, X., The information content of an open limit-order book. {\itshape Journal of Futures Markets: Futures, Options, and Other Derivative Products}, 2009, \textbf{29}, 16--41.

\bibitem[\protect\citeauthoryear{Chakraborty and Kearns}{2011}]{chakraborty2011market}
Chakraborty, T. and Kearns, M., Market making and mean reversion. In {\itshape Proceedings of the }{\itshape Proceedings of the 12th ACM conference on Electronic commerce}, pp. 307--314, 2011.

\bibitem[\protect\citeauthoryear{Challet {\itshape{et~al.}}}{2018}]{challet2018statistically}
Challet, D., Chicheportiche, R., Lallouache, M. and Kassibrakis, S., Statistically validated lead-lag networks and inventory prediction in the foreign exchange market. {\itshape Advances in Complex Systems}, 2018, \textbf{21}, 1850019.

\bibitem[\protect\citeauthoryear{Cont}{2001}]{cont2001empirical}
Cont, R., Empirical properties of asset returns: stylized facts and statistical issues. {\itshape Quantitative finance}, 2001, \textbf{1}, 223.

\bibitem[\protect\citeauthoryear{Cont {\itshape{et~al.}}}{2023}]{cont2023analysis}
Cont, R., Cucuringu, M., Glukhov, V. and Prenzel, F., Analysis and modeling of client order flow in limit order markets. {\itshape Quantitative Finance}, 2023, \textbf{23}, 187--205.

\bibitem[\protect\citeauthoryear{De~Long {\itshape{et~al.}}}{1990}]{delong1990noise}
De~Long, J.B., Shleifer, A., Summers, L.H. and Waldmann, R.J., Noise Trader Risk in Financial Markets. {\itshape Journal of Political Economy}, 1990.

\bibitem[\protect\citeauthoryear{DeLong {\itshape{et~al.}}}{1988}]{delong1988survival}
DeLong, J.B., Shleifer, A., Summers, L.H. and Waldmann, R.J., The survival of noise traders in financial markets. , 1988.

\bibitem[\protect\citeauthoryear{Farmer and Foley}{2009}]{farmer2009economy}
Farmer, J.D. and Foley, D., The economy needs agent-based modelling. {\itshape Nature}, 2009, \textbf{460}, 685--686.

\bibitem[\protect\citeauthoryear{Fischer and Krauss}{2018}]{fischer2018deep}
Fischer, T. and Krauss, C., Deep Learning for Price Prediction in Financial Markets. {\itshape European Journal of Operational Research}, 2018.

\bibitem[\protect\citeauthoryear{Gould {\itshape{et~al.}}}{2013}]{gould2013limit}
Gould, M.D., Porter, M.A., Williams, S., McDonald, M., Fenn, D.J. and Howison, S.D., Limit order books. {\itshape Quantitative Finance}, 2013, \textbf{13}, 1709--1742.

\bibitem[\protect\citeauthoryear{Grossman and Miller}{1988}]{grossman1988liquidity}
Grossman, S.J. and Miller, M.H., Liquidity and Market Structure. {\itshape Journal of Finance}, 1988.

\bibitem[\protect\citeauthoryear{Guyon and Elisseeff}{2003}]{guyon2003introduction}
Guyon, I. and Elisseeff, A., An introduction to variable and feature selection. {\itshape Journal of machine learning research}, 2003, \textbf{3}, 1157--1182.

\bibitem[\protect\citeauthoryear{Hagstr{\"o}mer and Nord{\'e}n}{2013}]{hagstromer2013diversity}
Hagstr{\"o}mer, B. and Nord{\'e}n, L., The diversity of high-frequency traders. {\itshape Journal of Financial Markets}, 2013, \textbf{16}, 741--770.

\bibitem[\protect\citeauthoryear{Huberman and Stanzl}{2005}]{huberman2005optimal}
Huberman, G. and Stanzl, W., Optimal liquidity trading. {\itshape Review of finance}, 2005, \textbf{9}, 165--200.

\bibitem[\protect\citeauthoryear{Kalay and Wohl}{2009}]{kalay2009detecting}
Kalay, A. and Wohl, A., Detecting liquidity traders. {\itshape Journal of Financial and Quantitative Analysis}, 2009, \textbf{44}, 29--54.

\bibitem[\protect\citeauthoryear{Kirilenko {\itshape{et~al.}}}{2017}]{kirilenko2017flash}
Kirilenko, A., Kyle, A.S., Samadi, M. and Tuzun, T., The flash crash: High-frequency trading in an electronic market. {\itshape The Journal of Finance}, 2017, \textbf{72}, 967--998.

\bibitem[\protect\citeauthoryear{Liaw {\itshape{et~al.}}}{2018}]{liaw2018tune}
Liaw, R., Liang, E., Nishihara, R., Moritz, P., Gonzalez, J.E. and Stoica, I., Tune: A research platform for distributed model selection and training. {\itshape arXiv preprint arXiv:1807.05118}, 2018.

\bibitem[\protect\citeauthoryear{Liu {\itshape{et~al.}}}{2017}]{liu2017survey}
Liu, W., Wang, Z., Liu, X., Zeng, N., Liu, Y. and Alsaadi, F.E., A survey of deep neural network architectures and their applications. {\itshape Neurocomputing}, 2017, \textbf{234}, 11--26.

\bibitem[\protect\citeauthoryear{Lo {\itshape{et~al.}}}{2000}]{lo2000foundations}
Lo, A.W., Mamaysky, H. and Wang, J., Foundations of Technical Analysis. {\itshape Journal of Finance}, 2000.

\bibitem[\protect\citeauthoryear{M{\"a}kinen {\itshape{et~al.}}}{2019}]{makinen2019forecasting}
M{\"a}kinen, Y., Kanniainen, J., Gabbouj, M. and Iosifidis, A., Forecasting jump arrivals in stock prices: new attention-based network architecture using limit order book data. {\itshape Quantitative Finance}, 2019, \textbf{19}, 2033--2050.

\bibitem[\protect\citeauthoryear{Menkveld}{2013}]{menkveld2013high}
Menkveld, A.J., High frequency trading and the new market makers. {\itshape Journal of financial Markets}, 2013, \textbf{16}, 712--740.

\bibitem[\protect\citeauthoryear{Montavon {\itshape{et~al.}}}{2019}]{montavon2019layer}
Montavon, G., Binder, A., Lapuschkin, S., Samek, W. and M{\"u}ller, K.R., Layer-wise relevance propagation: an overview. {\itshape Explainable AI: interpreting, explaining and visualizing deep learning}, 2019, pp. 193--209.

\bibitem[\protect\citeauthoryear{Mourer {\itshape{et~al.}}}{2023}]{mourer2023selecting}
Mourer, A., Forest, F., Lebbah, M., Azzag, H. and Lacaille, J., Selecting the number of clusters k with a stability trade-off: an internal validation criterion. In {\itshape Proceedings of the }{\itshape Pacific-Asia Conference on Knowledge Discovery and Data Mining}, pp. 210--222, 2023.

\bibitem[\protect\citeauthoryear{Powers}{2011}]{powers2011evaluation}
Powers, D., Evaluation: From Precision, Recall and F-Measure to ROC, Informedness, Markedness \& Correlation. {\itshape Journal of Machine Learning Technologies}, 2011, \textbf{2}, 37--63.

\bibitem[\protect\citeauthoryear{Preis}{2011}]{preis2011price}
Preis, T., Price-time priority and pro rata matching in an order book model of financial markets. In {\itshape Econophysics Of Order-driven Markets: Proceedings Of Econophys-Kolkata V}, pp. 65--72, 2011, Springer.

\bibitem[\protect\citeauthoryear{Shleifer and Summers}{1990}]{shleifer1990noise}
Shleifer, A. and Summers, L.H., The noise trader approach to finance. {\itshape Journal of Economic perspectives}, 1990, \textbf{4}, 19--33.

\bibitem[\protect\citeauthoryear{Staccioli and Napoletano}{2021}]{staccioli2021agent}
Staccioli, J. and Napoletano, M., An agent-based model of intra-day financial markets dynamics. {\itshape Journal of Economic Behavior \& Organization}, 2021, \textbf{182}, 331--348.

\bibitem[\protect\citeauthoryear{Van~Kervel and Menkveld}{2019}]{van2019high}
Van~Kervel, V. and Menkveld, A.J., High-frequency trading around large institutional orders. {\itshape The Journal of Finance}, 2019, \textbf{74}, 1091--1137.

\bibitem[\protect\citeauthoryear{Vyetrenko {\itshape{et~al.}}}{2020}]{vyetrenko2020get}
Vyetrenko, S., Byrd, D., Petosa, N., Mahfouz, M., Dervovic, D., Veloso, M. and Balch, T., Get real: Realism metrics for robust limit order book market simulations. In {\itshape Proceedings of the }{\itshape Proceedings of the First ACM International Conference on AI in Finance}, pp. 1--8, 2020.

\bibitem[\protect\citeauthoryear{Wah {\itshape{et~al.}}}{2017}]{wah2017welfare}
Wah, E., Wright, M. and Wellman, M.P., Welfare effects of market making in continuous double auctions. {\itshape Journal of Artificial Intelligence Research}, 2017, \textbf{59}, 613--650.

\bibitem[\protect\citeauthoryear{Wilinski}{2025}]{wilinski2025simulator}
Wilinski, M., https://github.com/mateuszwilinski/lob-abm-simulator/. , 2025.

\bibitem[\protect\citeauthoryear{Wilinski and Goel}{2025}]{wilinski2025classifying}
Wilinski, M. and Goel, A., https://github.com/mateuszwilinski/cluster-investors/. , 2025.

\end{thebibliography}

\end{document}